\documentclass[aps,prd,onecolumn,eqsecnum, nofootinbib]{revtex4-2}
\usepackage{amssymb}
\usepackage{amsmath}
\usepackage{amsfonts}
\usepackage{graphicx}
\usepackage{bm}
\usepackage{hyperref}
\DeclareSymbolFont{EulerScript}{U}{eus}{m}{n}
\SetSymbolFont{EulerScript}{bold}{U}{eus}{b}{n}
\DeclareSymbolFontAlphabet\scrpt{EulerScript}
\newcommand{\E}{{\cal E}} 
\newcommand{\Eq}{{\cal E}^{\sf q}} 
 
\newcommand{\Eh}{{\cal E}^{\sf h}} 
 
\newcommand{\EEm}{({\cal EE})^{\sf m}} 
\newcommand{\EEq}{({\cal EE})^{\sf q}} 
\newcommand{\EEh}{({\cal EE})^{\sf h}}
\newcommand{\ovr}[3]{{\mbox{}_{#2}{#1}_{{\sf #3}}}} 
\newcommand{\e}[1]{{$\times 10^{#1}$}}
\allowdisplaybreaks
\begin{document}
\title{Impact of nonlinearities on relativistic dynamical tides in compact binary inspirals}  
\author{Tristan Pitre}
\email{tpitre@uoguelph.ca} 
\author{Eric Poisson}
\email{epoisson@uoguelph.ca}
\affiliation{Department of Physics, University of Guelph, Guelph, Ontario, N1G 2W1, Canada} 
\date{June 9, 2025} 
\begin{abstract}
The tidal deformation of a neutron star in a binary inspiral driven by the emission of gravitational waves affects the orbital dynamics and produces a measurable modulation of the waves. Late in the inspiral, the external timescale associated with changes in the tidal field becomes comparable to internal timescales associated with hydrodynamical processes within the star, and a regime of dynamical tides takes over from a prior regime of static tides. A recent analysis by Yu {\it et al.}\
[M.N.R.A.S. {\bf 519}, 4325 (2022)] reveals that nonlinear aspects of the tidal interaction are important during the regime of dynamical tides. Their theoretical framework is grounded in Newtonian gravity and fluid mechanics, and relies on a representation of the tidal deformation in terms of the star's normal modes of vibration; the nonlinearities are incorporated through a coupling of the modes and a nonlinear driving by the external tidal field. We confirm their observation in a general relativistic treatment of the tidal deformation of a neutron star, without relying on a mode representation of this deformation. The starting point of our description is a simultaneous time-derivative and nonlinear expansion of the tidal deformation, expressed in terms of three encapsulating constants, the static $k_2$, dynamic $\ddot{k}_2$, and nonlinear $p_2$ tidal constants. We describe the neutron star's deformation in terms of a well-defined quadrupole moment tensor ${\cal Q}_{ab}$, which is related to the tidal quadrupole moment ${\cal E}_{ab}$ through a frequency-domain response function $\tilde{k}_2(\omega)$. In a pragmatic extension of our simultaneous time-derivative and nonlinear expansion, we express this in a form proportional to $(1-\omega^2/\omega_*^2)^{-1}$, the characteristic response of a harmonic oscillator subjected to a driving force of frequency $\omega$, with a natural-frequency parameter $\omega_*$ constructed from the tidal constants. We compute these for polytropic stellar models, and show that the nonlinear constant $p_2$ lowers the frequency parameter by as much as 15\% relative to an estimation based on a purely linear treatment of the tidal deformation. This means that the approach to resonance is triggered at an earlier frequency during an inspiral, which causes a significant enhancement of the star's tidal response.  
\end{abstract} 
\maketitle

\section{Introduction and summary} 
\label{sec:intro} 

\subsection{Dynamical tides in binary inspirals} 

A neutron star in a binary inspiral driven by the emission of gravitational waves gets progressively deformed by the tidal forces exerted by its companion, and the tidal deformation produces a subtle modulation of the gravitational-wave signal \cite{flanagan-hinderer:08}. A measurement of this effect discloses important information about the equation of state of the extremely dense matter inside a neutron star, which is currently poorly constrained \cite{ozel-freire:16, oertel-etal:17, baym-etal:18}. Such a measurement was attempted for the gravitational-wave event GW170817 \cite{GW170817:17, GW170817:18, narikawa-uchikata-tanaka:21}, which famously featured the merger of two neutron stars in a multimessenger event. While the measurement could only reveal an upper bound for the tidal deformability of a neutron star, it was nevertheless informative, as it favors a relatively soft equation of state that produces a relatively small neutron star \cite{landry-essick-reed-chatziioannou:20}. These exiting developments are reviewed in Ref.~\cite{chatziioannou:20}, and prospects for future measurements are summarized in Ref.~\cite{pacilio-maselli-fasano-pani:22}.

The physics of tides in a binary inspiral can be classified into regimes of {\it static} and {\it dynamical} tides. In the regime of static tides, which occurs early in the inspiral, the external timescale associated with changes in the tidal field --- the binary's orbital period --- is very long compared with internal timescales associated with hydrodynamical processes taking place within the neutron star; in this regime the tidal field can be idealized as static. In the regime of dynamical tides, which occurs later in the inspiral, all timescales become comparable, and the time dependence of the tidal field becomes important. It was established \cite{hinderer-etal:16, steinhoff-etal:16} that dynamical tides play a significant role in the emission of gravitational waves, and that they will soon be accessible to measurement. This realization led to a thorough examination of dynamical tides in binary inspirals, including the formulation of models of a dynamically deformed neutron star \cite{steinhoff-etal:21, andersson-pnigouras:21, passamonti-andersson-pnigouras:21, passamonti-andersson-pnigouras:22}, calculations of gravitational-wave signals \cite{schmidt-hinderer:19, mandal-etal:23, mandal-etal:24}, and the identification of what can be learned from them \cite{andersson-ho:18, williams-pratten-schmidt:22, pratten-schmidt-williams:22}.

The main motivation for the work presented here is the recent observation by Yu {\it et al.}\ that the regime of dynamical tides is strongly influenced by nonlinear aspects of the tidal dynamics \cite{yu-etal:22}. These authors showed that nonlinearities can lead to a phase shift in the gravitational-wave signal that amounts to a correction of order 10\% -- 20\% to the prediction of the linearized description. They arrived at this result by adopting a Newtonian description of the fluid mechanics and gravity inside a neutron star, by representing the tidal deformation in terms of the star's normal modes of vibration, and by examining the nonlinear driving and coupling of these modes. (The theoretical framework behind these calculations was provided by Ref.~\cite{weinberg-etal:12}, based on a foundation given in Ref.~\cite{schenk-etal:01}.) The dominant effect comes from an effective shift in the natural frequencies of the implicated modes. Our purpose with this paper is to confirm the conclusion of Yu {\it et al.}\ in a description of the tidal dynamics that is firmly grounded in general relativity, without relying on a mode representation of the tidal deformation. Other approaches to this problem were recently pursued by Nouri {\it et al.}\ in relativistic perturbation theory \cite{nouri-etal:22} and by Kuan, Kiuchi, and Shibata in numerical relativity \cite{kuan-kiuchi-shibata:24}. 

There are other situations in which nonlinear aspects of the tidal dynamics of a binary inspiral might be important. A prominent example is an instability that results from the nonlinear coupling between a neutron star's $p$-modes and $g$-modes \cite{weinberg-etal:13, weinberg:16}. The prospect of observing this effect in gravitational-wave signals was explored in Refs.~\cite{essick-vitale-weinberg:16, essick-weinberg:18}, and whether it was actually detected in GW170817 was debated \cite{GW170817:19, reyes-brown:20}. An intriguing possibility is that nonlinearities could produce a resonant locking of a $g$-mode frequency with the frequency of the tidal field \cite{kwon-yu-venumadhav:24}. The work described below, however, has nothing to do with $p$-modes or $g$-modes. If modes are to be associated with our results, the only ones that are implicated are the star's fundamental modes.\footnote{We use the plural because an $f$-mode is defined for each value of the azimuthal index $m$. For a linear tidal deformation dominated by its quadrupole ($\ell=2$) contribution, the relevant values are $m = \{-2, 0, 2\}$. Modes with $m = \pm 1$ do not couple to a quadrupolar tidal field.} 

Although the mode picture will be suggested in our description of the tidal response function of a neutron star (see Sec.~\ref{intro:response} below), we insist that our relativistic framework does not rely on a mode representation of the tidal deformation. Our reasons to stay away from modes in general relativity were explained at length in Sec.~I C of our earlier work on dynamical tides \cite{pitre-poisson:24}. We shall not rehearse these arguments here, but merely iterate that a mode representation is problematic in general relativity (at least in the current stage of developments). Fortunately we have access to an alternative approach to the relativistic description of tides, as developed in Ref.~\cite{poisson:21a}, and we shall make extensive use of it in this paper. 

\subsection{Tidal deformation of a body: Newtonian theory}
\label{intro:newton}

To describe the approach to dynamical tides proposed in Ref.~\cite{poisson:21a}, we consider a star of mass $M$ and radius $R$ immersed in a tidal environment created by a companion, and first examine the situation within the framework of the Newtonian theory of gravitation. In the next subsection we will port the discussion to general relativity and see that the main ingredients of the description are preserved in the promotion to strong fields.  

In a Newtonian context, the gravitational potential outside the deformed body can be expressed as
\begin{align} 
U_{\rm out} &= \frac{GM}{r} - \frac{1}{2} \biggl( r^2
+ 2k_2 \frac{R^5}{r^3} \biggr)\, {\cal E}_{ab} \Omega^a \Omega^b 
+ \ddot{k}_2 \frac{R^8}{GM} \frac{1}{r^3}\, \ddot{\cal E}_{ab} \Omega^a \Omega^b
+ p_2 \frac{R^8}{GM} \frac{1}{r^3}\, {\cal E}_{c\langle a} {\cal E}^c_{\ b\rangle}\Omega^a \Omega^b  
+ \cdots, 
\label{U_out} 
\end{align} 
in terms of a set of tidal constants $\{k_2, \ddot{k}_2, p_2\}$ that encapsulate the body's deformation.\footnote{These quantities are often called Love numbers, but we refrain from using this terminology in this paper. The reason is that too many distinct things have been called Love numbers in the literature, and this tends to lead to confusion.} Here, $G$ is the gravitational constant and $r$ is the distance from the star's center of mass. The tidal quadrupole moment ${\cal E}_{ab}$ --- a spatially constant, symmetric-tracefree, Cartesian tensor --- is defined by 
\begin{equation}
\E_{ab}(t) := -\partial_{ab} U^{\rm ext} \Bigr|_{\bm{x}=\bm{0}} 
\label{E_def} 
\end{equation}
in which $U^{\rm ext}$ is the external potential created by the companion object, and the right-hand side is evaluated at the body's center of mass, which is placed at the origin of the coordinate system. Overdots on ${\cal E}_{ab}$ (but not on $k_2$ --- $\ddot{k}_2$ is a constant independent of $k_2$) indicate differentiation with respect to time, and the angular brackets around tensorial indices are an instruction to symmetrize all indices and remove all traces.\footnote{For example,  $A_{\langle ab \rangle} := A_{(ab)} - \frac{1}{3} \delta_{ab} A^c_{\ c}$ for any tensor $A_{ab}$, with $A_{(ab)} := \frac{1}{2}( A_{ab} + A_{ba})$.} Finally,
\begin{equation}
\Omega^a := [\sin\theta\cos\phi, \sin\theta\sin\phi, \cos\theta]
\label{Omega_def} 
\end{equation}
is the radial unit vector, which points away from the body's center of mass and is parametrized by the polar angles $(\theta,\phi)$. The combination $A_{\langle ab \rangle} \Omega^a \Omega^b$ for any spatially constant, symmetric-tracefree tensor $A_{\langle ab \rangle}$ is a spherical harmonic of degree $\ell = 2$; in Eq.~(\ref{U_out}), these spherical harmonics are multiplied by appropriate powers of $r$ so that each term is separately a solution to Laplace's equation. 

The gravitational potential of Eq.~(\ref{U_out}) features a growing term proportional to $r^2$; this is the tidal potential created by the companion object. It features also decaying terms proportional to $r^{-3}$; these describe the body's response to the applied tidal field. The term proportional to $k_2$ represents the static and linear response, the one involving $\ddot{k}_2$ describes the dynamic and linear response, and the term with $p_2$ represents the star's quadratic response. We note that the $\ddot{k}_2$ and $p_2$ inserted in Eq.~(\ref{U_out}) are defined here with a relative minus sign compared with the definitions proposed in Ref.~\cite{poisson:21a}; this ensures that their numerical value is positive instead of negative.

The potential of Eq.~(\ref{U_out}) incorporates only the quadrupole piece of the tidal deformation, and it ignores higher-order multipole moments. A more accurate description of the tidal environment would include higher-order moments (involving additional derivatives of the external potential and higher powers of $r$), and these would induce, at the linear level, a response that also includes higher-order moments (with additional tidal constants). At the quadratic level, the coupling of ${\cal E}_{ab}$ with itself necessarily generates monopole ($\ell = 0$) and hexadecapole ($\ell = 4$) terms in addition to the quadrupole ($\ell=2$) term featured in Eq.~(\ref{U_out}); these are omitted from Eq.~(\ref{U_out}) to keep the discussion to its essentials. In addition, the coupling of ${\cal E}_{ab}$ with higher-order tidal moments also generates additional multipoles; many of those were included in the computations of Ref.~\cite{poisson:21a}, but they are not considered here.   

The description of the star's tidal deformation contained in Eq.~(\ref{U_out}) is based on two assumptions that play a key role in our developments in this paper. The first is that the tidal interaction is weak, so that the deformation is small. In this context, the linear terms proportional to ${\cal E}_{ab}$ are taken to be small, and they dominate the deformation; the quadratic term proportional to ${\cal E}_{c\langle a} {\cal E}^c_{\ b\rangle}$ is a correction of the second order. The second assumption is that the timescale associated with changes in the tidal field --- the orbital period --- is long compared with timescales --- of order of magnitude comparable to $(R^3/GM)^{1/2}$ --- associated with hydrodynamical processes taking place inside the star. In this context, the term proportional to $\ddot{\cal E}_{ab}$ is small compared with the one involving the undifferentiated tidal moment, and terms involving additional derivatives are neglected. Taken together, these assumptions allow us to express $U_{\rm out}$ as a simultaneous nonlinear and time-derivative expansion, the first in powers of $U^{\rm ext}$, the second in the number of time derivatives. One of our central messages below is the observation that the physical effects associated with $\ddot{k}_2$ and $p_2$ are of the same order of magnitude: {\it dynamical and nonlinear aspects of the tidal interaction necessarily come together}. 

\subsection{Tidal deformation of a body: General relativity}
\label{intro:GR}

The relativistic analogue of Eq.~(\ref{U_out}) is the spacetime metric $g_{\alpha\beta}$ of a tidally deformed body, also computed in a simultaneous nonlinear and time-derivative expansion. This is constructed in Ref.~\cite{poisson:21a}, and a representative sample of the final result is (we now set $G=1$) 
\begin{align} 
g_{tt} &= -1 + \frac{2M}{r} 
- \biggl[ r^2 (1 + \cdots) + 2 k_2 \frac{R^5}{r^3} (1 + \cdots) \biggr]\,
{\cal E}_{ab} \Omega^a \Omega^b 
\nonumber \\ & \quad \mbox{}
- \biggl[ \frac{11}{42} r^4 (1 + \cdots) + k_2 \frac{R^5}{r} (1 + \cdots)
+ 2 \ddot{k}_2 \frac{R^8}{M} \frac{1}{r^3} (1 + \cdots) \biggr]\,
\ddot{\cal E}_{ab} \Omega^a \Omega^b. 
\nonumber \\ & \quad \mbox{}
- \biggl[ \frac{2}{7} r^4 (1 + \cdots)  + \frac{8}{7} k_2 \frac{R^5}{r} (1 + \cdots)
+ \frac{8}{7} k_2^2 \frac{R^{10}}{r^6} (1 + \cdots) 
+ 2 p_2 \frac{R^8}{M} \frac{1}{r^3} (1 + \cdots) \biggr]\,
{\cal E}_{c\langle a} \E^c_{\ b\rangle} \Omega^a \Omega^b.  
\label{metric_sample} 
\end{align} 
We recognize the same ingredients that appear in the Newtonian potential of Eq.~(\ref{U_out}): the tidal quadrupole moment ${\cal E}_{ab}(t)$ in its undifferentiated, differentiated, and squared forms, as well as the tidal constants $k_2$, $\ddot{k}_2$, and $p_2$. In this relativistic context, ${\cal E}_{ab}$ is no longer related to an exterior potential; it is a function of coordinate time $t$ that stays undetermined when we integrate the Einstein field equations in a neighborhood of the deformed body. The tidal moment can be tied to external conditions by matching $g_{\alpha\beta}$ to another metric defined and calculated in a larger region of spacetime, as was carried out in Ref.~\cite{poisson:21a}.   

In Eq.~(\ref{metric_sample}), the factors denoted $(1 + \cdots)$ stand for functions of $M/r$ that are fully identified in Secs.~V C and V E of Ref.~\cite{poisson:21a}, with explicit expressions provided in Apps.~C and D of this reference; these functions all admit an expansion in powers of $M/r$ that behaves as $1 + O(M/r)$. We observe that the description of the tidal field is more complicated in general relativity than in the Newtonian theory, with time-derivative and quadratic contributions proportional to $r^4$. The description of the tidal response is also more complicated, with $k_2$ appearing in the time-derivative and quadratic terms, in addition to the expected $\ddot{k}_2$ and $p_2$. As with the Newtonian potential of Eq.~(\ref{U_out}), the metric of Eq.~(\ref{metric_sample}) focuses entirely on the quadrupole ($\ell = 2$) piece of the tidal deformation, and it omits monopole ($\ell = 0$) and hexadecapole ($\ell = 4$) terms that also result from the coupling of ${\cal E}_{ab}$ with itself. These additional pieces, however, will be included in our technical developments below. 

\subsection{This work: Computation of $p_2$}
\label{intro:p2}

The static tidal constant $k_2$ was calculated by many researchers for a broad range of stellar models (see, for example, Fig.~1 of Ref.~\cite{chatziioannou:20}). In Ref.~\cite{pitre-poisson:24} we computed the dynamic tidal constant $\ddot{k}_2$ for relativistic polytropes, and explained its relevance to a description of dynamical tides in a binary inspiral. Our main goal in this paper is to compute the quadratic tidal constant $p_2$, also for relativistic polytropes, and to explain its importance in an improved description of dynamical tides.

\begin{figure}
\includegraphics[width=0.6\linewidth]{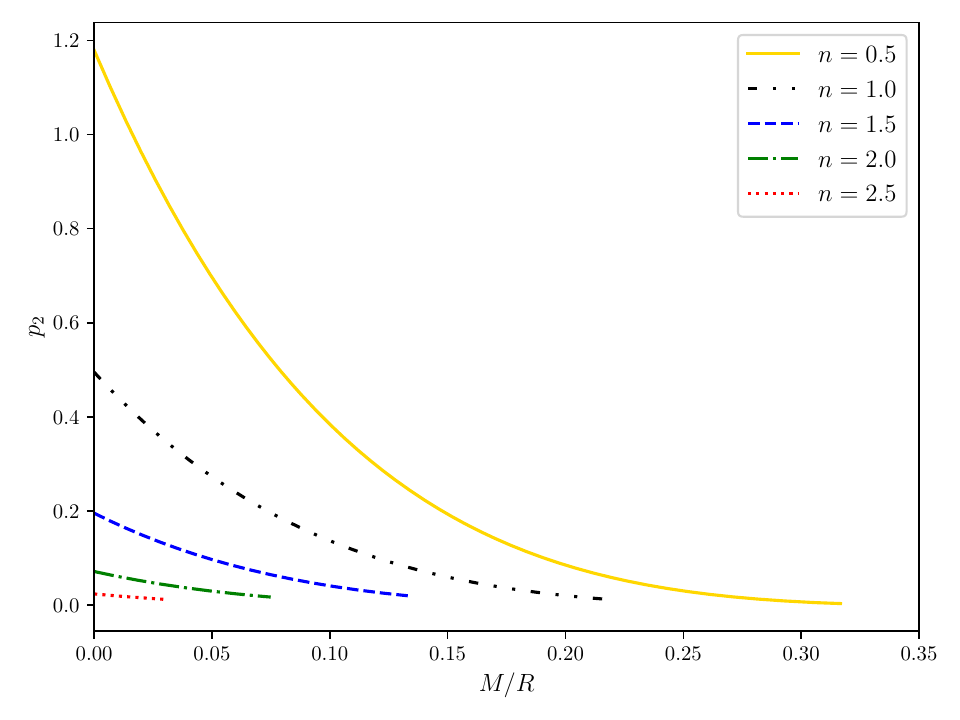}
\caption{Quadratic tidal constant $p_2$ as a function of stellar compactness $M/R$. The computation is carried out for polytropic stellar models with equation of state $p = K \rho^{1+1/n}$. Yellow curve: $n=0.5$. Black curve: $n=1$. Blue curve: $n=1.5$. Green curve: $n=2.0$. Red curve: $n=2.5$.}  
\label{fig:p2} 
\end{figure} 

\begin{figure}
\includegraphics[width=0.49\linewidth]{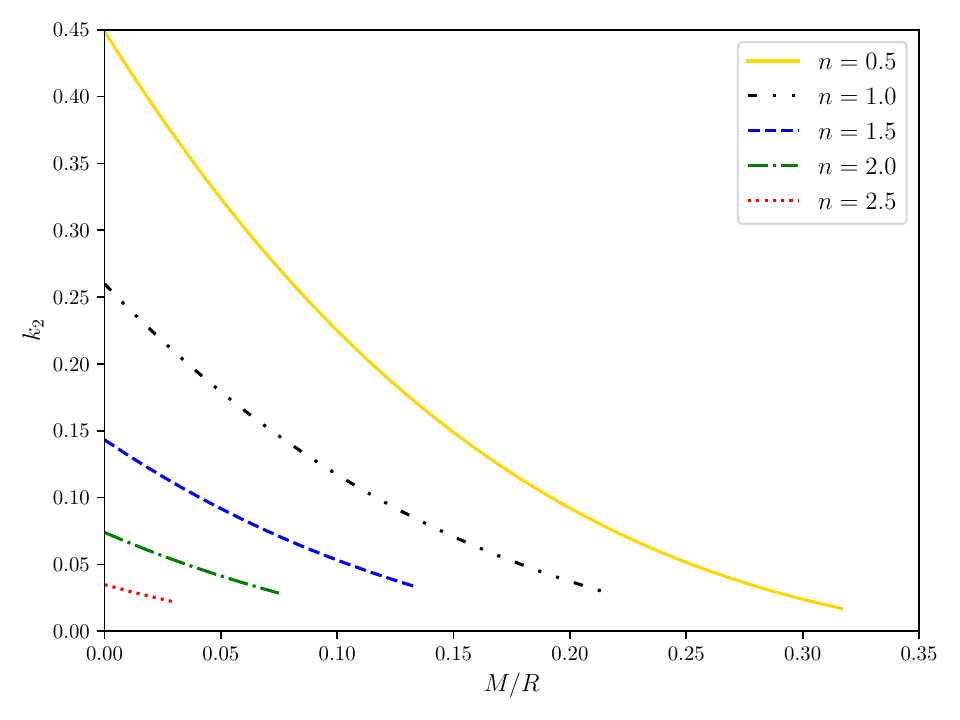}
\includegraphics[width=0.49\linewidth]{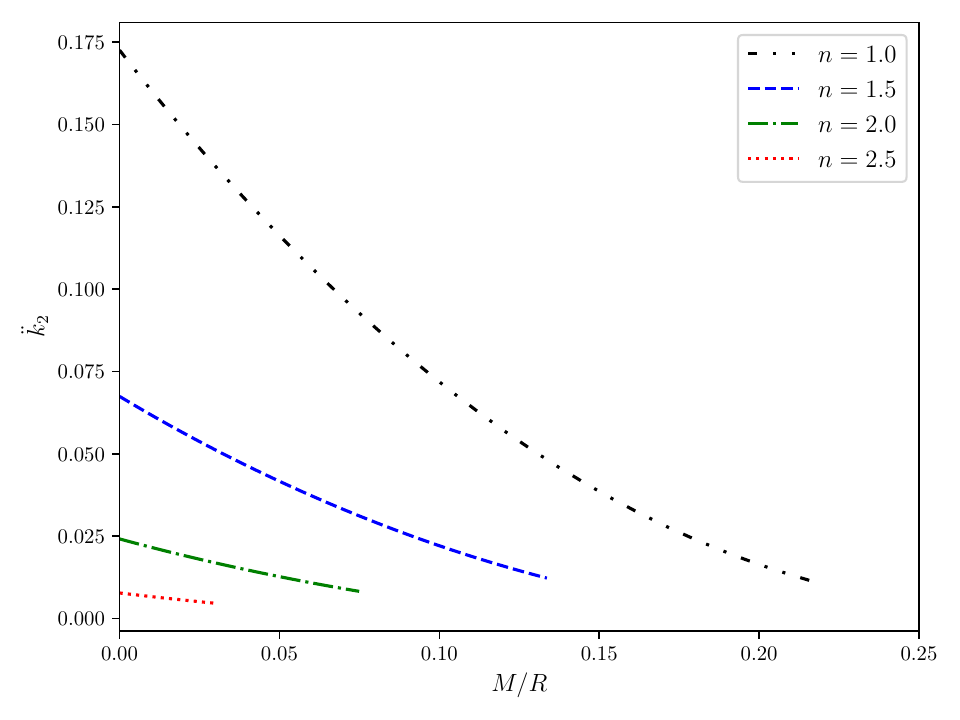}
\caption{Static and dynamic tidal constants as functions of stellar compactness $M/R$, computed for polytropes with equation of state $p = K \rho^{1+1/n}$. Left: Static constant $k_2$. Right: Dynamic constant $\ddot{k}_2$ (this figure is adapted from Ref.~\cite{pitre-poisson:24}). Yellow curve: $n=0.5$. Black curves: $n=1$. Blue curves: $n=1.5$. Green curves: $n=2.0$. Red curves: $n=2.5$.}  
\label{fig:k2} 
\end{figure} 

Our results for $p_2$ are displayed in Fig.~\ref{fig:p2}, where it is plotted as a function of compactness $M/R$ for polytropic stellar models with equation of state
\begin{equation} 
p = K \rho^{1+1/n},
\label{poly}
\end{equation}
where $p$ is the pressure, $\rho$ is the rest-mass density, $K$ is a constant, and $n$ is a constant polytropic index; we sample the values $n = \{ 0.5, 1.0, 1.5, 2.0, 2.5 \}$. The curves begin near the Newtonian limit of vanishing compactness and end at the configuration of maximum mass, at which the stellar models become dynamically unstable to radial perturbations. As is usual with tidal constants, $p_2$ decreases as $M/R$ increases for a given equation of state; for a given compactness, $p_2$ decreases as $n$ increases and the equation of state becomes softer (which produces a star that is more centrally dense).

To provide context for the results of Fig.~\ref{fig:p2}, we plot in Fig.~\ref{fig:k2} the linear constants $k_2$ and $\ddot{k}_2$ as functions of $M/R$ for the same polytropic models. (The results for the dynamic tidal constant are taken from Ref.~\cite{pitre-poisson:24}, which did not produce computations for $n=0.5$.) The figure reveals that $p_2$ decreases faster than $k_2$ as $M/R$ increases, and that the relative size of these constants depends on the equation of state. For example, for $n=1.0$, $p_2/k_2$ begins near 1.9 when $M/R \to 0$, and it decreases to approximately 0.5 when $M/R$ achieves its maximum value. For $n=2.0$ the limiting values are approximately 1.0 and 0.6, respectively. More importantly for our discussion below, we find that the ratio $p_2/\ddot{k}_2$ for a given compactness is approximately the same for any choice of $n$, starting near 2.9 when $M/R \to 0$ and decreasing steadily as $M/R$ increases. Because the maximum compactness decreases as $n$ increases (and the equation of state becomes softer), the limiting value of $p_2/\ddot{k}_2$ is observed to increase with $n$; for $n=1.0$ we have that the ratio is near 1.2 at maximum compactness, while it is approximately 2.1 for $n=2.0$.

\subsection{Tidal response: Induced quadrupole moment}
\label{intro:response}

The Newtonian potential of Eq.~(\ref{U_out}) reveals immediately that the quadrupole moment of the deformed mass distribution\footnote{This is defined by ${\cal Q}_{ab} := \int \rho (x_a x_b - \frac{1}{3} r^2 \delta_{ab})\, d^3x$, where $\rho$ is the mass density.} is given by (we now restore the gravitational constant) 
\begin{equation}
G {\cal Q}_{ab} = -\frac{2}{3} k_2 R^5\, {\cal E}_{ab}
+ \frac{2}{3} \ddot{k}_2 \frac{R^8}{GM}\, \ddot{\cal E}_{ab}
+ \frac{2}{3} p_2 \frac{R^8}{GM}\, {\cal E}_{c\langle a} {\cal E}^c_{\ b\rangle}
\label{Q_vs_E_intro}
\end{equation}
in relation to the tidal moment ${\cal E}_{ab}$. The relation applies also in general relativity and the metric of Eq.~(\ref{metric_sample}), but its origin and status are far more subtle than in the Newtonian context.

While the relation between $U_{\rm out}$ and the body's multipole structure is immediate in the Newtonian theory, in general relativity there is no native notion of multipole moments for an individual body in a system of bodies moving under a mutual gravitational attraction. Before a relation such as Eq.~(\ref{Q_vs_E_intro}) can be formulated in a relativistic setting, a definition for the body's multipole moments must be provided. An avenue for this was proposed in Ref.~\cite{poisson:21a}, in which each body is permitted to have an internal gravity that's arbitrarily strong (so that it must be described in full general relativity), but in which the mutual gravity between bodies is restricted to be weak (so that it can be described as a post-Newtonian approximation to general relativity). In this approach, the tidal constants $k_2$, $\ddot{k}_2$, and $p_2$ are creatures of general relativity, but the mass quadrupole moment ${\cal Q}_{ab}$ is a creature of post-Newtonian theory, and is defined only when the body interacts weakly with its companion. Note that this assumption is entirely compatible with those behind the simultaneous time-derivative and nonlinearity expansion introduced in Sec.~\ref{intro:newton}. The upshot of all this is that Eq.~(\ref{Q_vs_E_intro}) is valid in general relativity, to leading order in a post-Newtonian expansion of the system's mutual gravity, and exactly with regards to the body's strong self-gravity. Equation (\ref{Q_vs_E_intro}) implicates tidal constants computed in full general relativity, but it omits corrections of the first post-Newtonian order to ${\cal Q}_{ab}$ that were obtained in Ref.~\cite{poisson:21a}. 

Equation (\ref{Q_vs_E_intro}) provides a sound foundation to describe the tidal dynamics of a binary system in general relativity, and it gives an operational meaning to the tidal constants $k_2$, $\ddot{k}_2$, and $p_2$. In Sec.~\ref{sec:response} we show that the equation can be written in the frequency domain as 
\begin{equation} 
G \tilde{\cal Q}_{ab}(\omega) = -\frac{2}{3} \tilde{k}_2(\omega) R^5\, \tilde{\cal E}_{ab}(\omega),
\end{equation}
in terms of a tidal response function $\tilde{k}_2(\omega)$ given by 
\begin{equation}
\tilde{k}_2(\omega) = \frac{\tilde{k}_2(0)}{1 - \omega^2/\omega^2_*},
\label{ktilde3}
\end{equation}
where $\omega$ is the external frequency of the tidal field,
\begin{equation} 
\tilde{k}_2(0) := k_2
\end{equation}
is the static tidal constant, and
\begin{equation}
\omega_*^2 := \frac{k_2}{\ddot{k}_2 + \frac{1}{4} p_2 \frac{M'}{M+M'}}\, \frac{GM}{R^3}
\label{mode_frequency}
\end{equation}
is the square of a frequency parameter that encapsulates the body's dynamical response ($M'$ is the mass of the companion object). The right-hand side of Eq.~(\ref{ktilde3}) reproduces the characteristic behavior of a harmonic oscillator of natural frequency $\omega_*$ subjected to a driving force with frequency $\omega$.

Equation (\ref{ktilde3}) suggests that the tidal deformation can be represented in terms of the star's normal modes of vibration. In such a representation,\footnote{We argue at length in Sec.~I C of Ref.~\cite{pitre-poisson:24} that a mode representation of the tidal deformation is currently limited to a Newtonian description of the tidal dynamics.} each mode behaves as a driven harmonic oscillator, and the tidal response is expressed as an infinite sum over contributions from each mode. It is the case, however, that the mode sum is strongly dominated by a single term coming from the star's fundamental mode, and in this regime the tidal response takes the form of Eq.~(\ref{ktilde3}), with $\omega_*$ now denoting the $f$-mode frequency. We arrive at Eq.~(\ref{ktilde3}) via a very different route: we begin with a proper relativistic foundation based on Eq.~(\ref{Q_vs_E_intro}), we do not invoke a mode decomposition of the tidal deformation, and $\omega_*$ has nothing to do with a mode frequency. 

Equation (\ref{Q_vs_E_intro}) rests on our simultaneous time-derivative and nonlinearity expansion of the tidal interaction, in which $\omega$ is implicitly taken to be much smaller than $\omega_*$. In this context, it seems clear that the tidal response function should be expressed as
$\tilde{k}_2(\omega) \propto 1 + \omega^2/\omega_*^2$ instead of given by Eq.~(\ref{ktilde3}). This pragmatic repackaging of the response function is inspired by the mode representation of the tidal deformation, and it is expected to provide a faithful description of the star's dynamical response in a wider domain of validity in which $\omega$ is allowed to become comparable to $\omega_*$. This expectation was validated in Ref.~\cite{pitre-poisson:24}, where we showed that the frequency parameter $\omega_*$ is numerically very close to the star's $f$-mode frequency.\footnote{This comparison was done in a strictly linear description of the tidal deformation, so that $p_2$ was set equal to zero. We shall presently turn to the importance of including $p_2$ in the frequency parameter.}

Equation (\ref{ktilde3}) implies that just like a driven oscillator, a tidally deformed star undergoes a resonance when $\omega = \omega_*$. Because the frequency parameter scales with\footnote{In this equation, the factor of $2\pi$ is inserted to ease a conversion from rad/s for the angular frequency to Hz for the corresponding linear frequency.}
\begin{equation}
\sqrt{ \frac{GM}{R^3} } = (2\pi) (1,650) \biggl( \frac{M}{1.4\ M_\odot} \biggr)^{1/2}
\biggl( \frac{12\ \mbox{km}}{R} \biggr)^{3/2}\, \mbox{rad/s}, 
\label{freq_unit} 
\end{equation}
the resonance is never realized in the context of a binary inspiral, either because the merger occurs before $\omega$ reaches $\omega_*$, or because the resonant frequency lies outside the active band of current gravitational-wave detectors. Equation (\ref{ktilde3}), however, does an excellent job of capturing the approach to resonance, which manifests itself in a substantial amplification of $\tilde{k}_2(\omega)$ with respect to its static value. 

The main message of Eq.~(\ref{mode_frequency}) is that {\it the nonlinear aspects of the tidal interaction decrease the value of the frequency parameter, so that the approach to resonance occurs earlier during a binary inspiral}. This conclusion is in complete accord with Yu {\it et al.}\ \cite{yu-etal:22}, who calculated the impact of nonlinear tides on a binary inspiral within a Newtonian description of the tidal interaction. Their calculation was based on a mode representation of the tidal deformation, with each mode behaving as an anharmonic oscillator driven in a nonlinear fashion by the external tidal force. In their context, the frequency shift comes as a natural consequence of the nonlinear physics of the oscillators. Our computations provide an extension of their results to relativistic gravity, without the need to rely on modes. 

The mass fraction $M'/(M+M')$ that occurs next to $p_2$ in Eq.~(\ref{mode_frequency}) ranges from the small value $M'/M$ when $M' \ll M$ to a value of unity when $M' \gg M$; the nonlinear correction to the frequency parameter is larger when the companion has a larger mass.\footnote{This conclusion is based on the fiction that only the star of mass $M$ is undergoing a tidal deformation. The actual situation involves two deformed stars, and the optimum combination of masses is more subtle to determine.} In our considerations below we shall assume that the binary consists of equal-mass bodies, so that $M'/(M+M') = 1/2$. 

\begin{figure}
\includegraphics[width=0.6\linewidth]{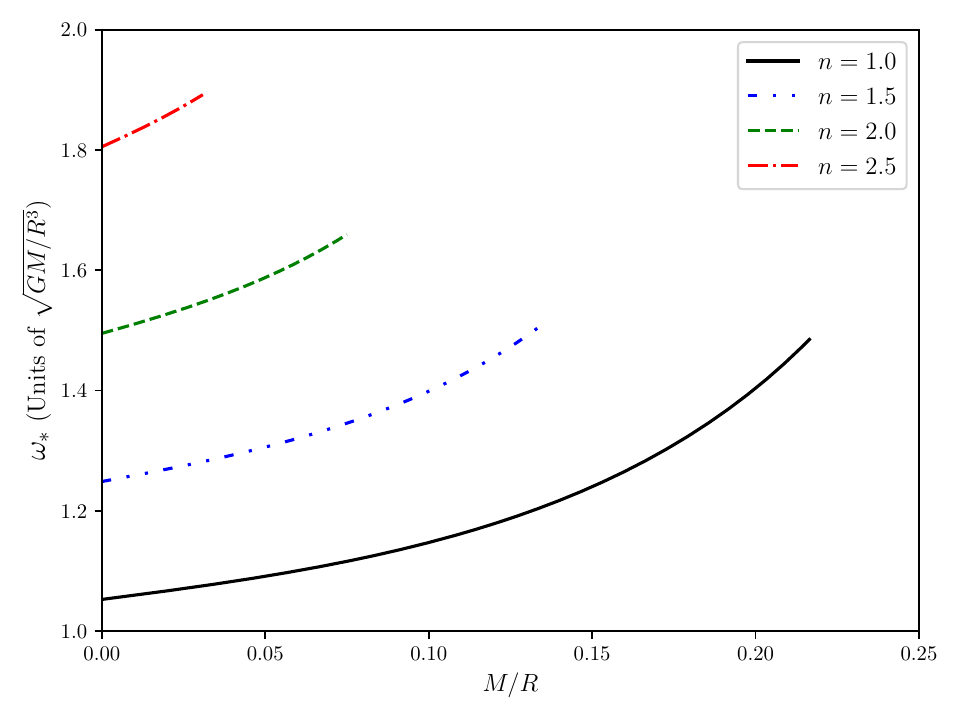}
\caption{Frequency parameter $\omega_*$ as a function of compactness $M/R$ for polytropic stellar models with different indices $n$, computed for $M' = M$. The frequency is given in units of $(GM/R^3)^{1/2}$. Each curve begins near the Newtonian limit $M/R = 0$ and ends at the configuration of maximum mass, at which the stellar model becomes dynamically unstable to radial perturbations. Black curve: $n=1.0$. Blue curve: $n=1.5$. Green curve: $n=2.0$. Red curve: $n=2.5$.}
\label{fig:frequency} 
\end{figure} 

\begin{figure}
\includegraphics[width=0.6\linewidth]{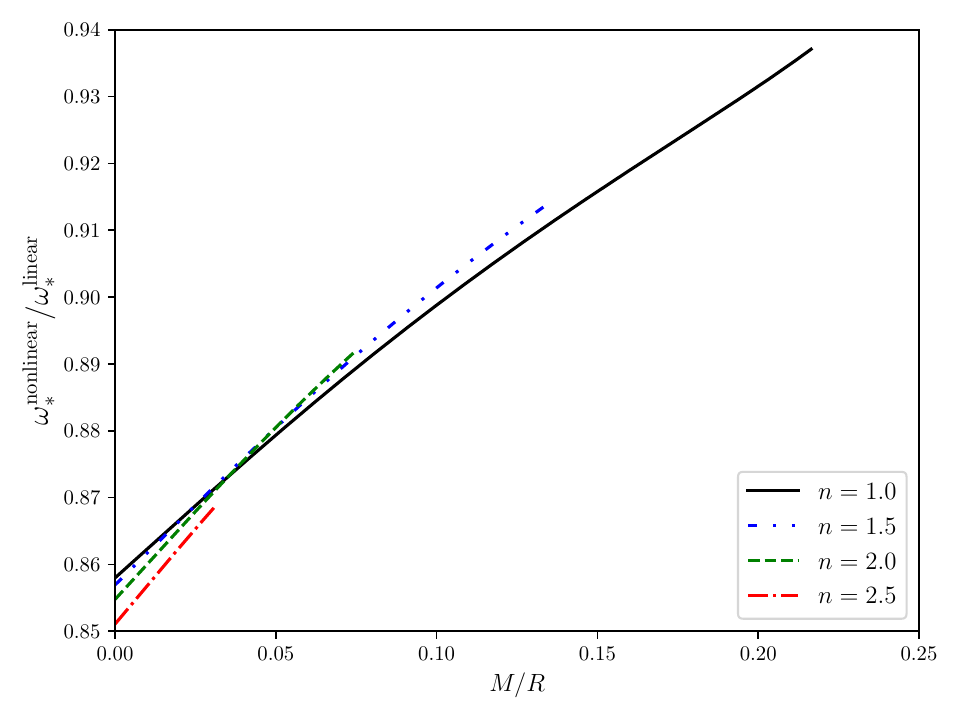}
\caption{Ratio of nonlinear to linear frequency parameters as a function of compactness $M/R$ for polytropic stellar models with different indices $n$, computed for $M' = M$. Black curve: $n=1.0$. Blue curve: $n=1.5$. Green curve: $n=2.0$. Red curve: $n=2.5$.} 
\label{fig:frequency_ratio} 
\end{figure} 

In Fig.~\ref{fig:frequency} we plot the frequency parameter $\omega_*$ as a function of stellar compactness $M/R$ for polytropic stellar models based on Eq.~(\ref{poly}). The frequency is presented in units of $(GM/R^3)^{1/2}$, and in those $\omega_*$ is of order unity. For a given polytropic index $n$, the frequency increases moderately (by approximately $20\%$) as the compactness increases to its maximum value. For a given compactness, $\omega_*$ increases with $n$; a softer equation of state (larger $n$) gives rise to a larger frequency parameter.

In Fig.~\ref{fig:frequency_ratio} we plot the ratio $\omega_*^{\rm nonlinear}/\omega_*^{\rm linear}$ as a function of $M/R$, in which $\omega_*^{\rm nonlinear}$ is the nonlinear expression for the frequency parameter given by Eq.~(\ref{mode_frequency}), while $\omega_*^{\rm linear}$ is the linear approximation calculated with $p_2$ set to zero. The plot reveals that irrespective of the equation of state, the ratio is approximately $0.85$ for a star with small compactness, and rises to approximately $0.94$ for a large compactness; for a given mass, the impact of nonlinear tides is greater for large stars. That $\omega_*^{\rm nonlinear}$ can deviate substantially from $\omega_*^{\rm linear}$ has to do with the large numerical value of $p_2$ compared with $\ddot{k}_2$; this can be seen from Figs.~\ref{fig:p2} and \ref{fig:k2}, as well as Table~\ref{tab:love} in Sec.~\ref{sec:love} below. This substantial deviation is the central result of this work; it shows that nonlinearities make an essential contribution to dynamical tides. 

\begin{figure}
\includegraphics[width=0.6\linewidth]{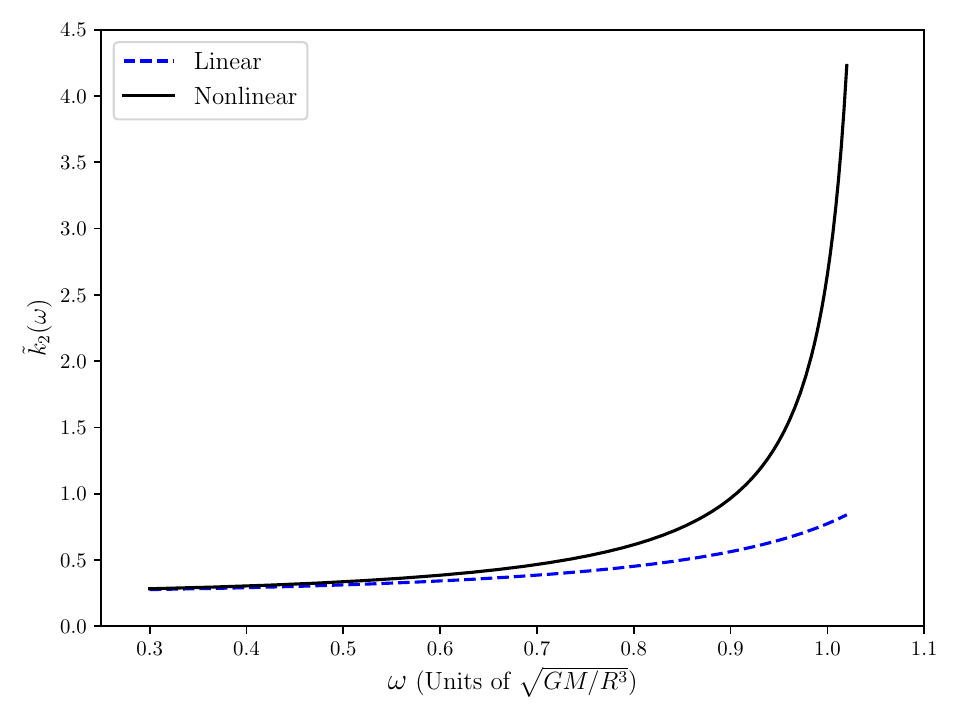}
\caption{Plot of $\tilde{k}_2(\omega)$ as a function of frequency $\omega$ for an $n=1$ polytrope with a small compactness $M/R$. The external frequency $\omega$ is presented in units of $(GM/R^3)^{1/2}$. Black curve: tidal response with $\omega_* = \omega_*^{\rm nonlinear}$. Blue curve: tidal response with $\omega_* = \omega_*^{\rm linear}$. The tidal response function was computed for $M/R = 1.9999 \times 10^{-5}$, which gives rise to $k_2 = 0.259872$, $\ddot{k}_2 = 0.172580$, $p_2 = 0.494857$, $M = M'$, so that $\omega_*^{\rm nonlinear} = 1.05285$ while $\omega_*^{\rm linear} = 1.22711$.} 
\label{fig:tidal_response} 
\end{figure} 

In Fig.~\ref{fig:tidal_response} we illustrate the impact of the nonlinear correction to the frequency parameter on the tidal response of a compact star. We plot two versions of $\tilde{k}_2(\omega)$, as given by Eq.~(\ref{ktilde3}). The first is evaluated with $\omega_*$ set equal to $\omega_*^{\rm nonlinear}$, the nonlinear expression of Eq.~(\ref{mode_frequency}). The second is evaluated with $\omega_* = \omega_*^{\rm linear}$, in which $p_2$ is set to zero. The figure shows vividly that the smaller value of $\omega_*^{\rm nonlinear}$ can have a dramatic effect on the tidal response: the earlier onset of resonance as $\omega$ increases toward $\omega_*$ produces a strong magnification of $\tilde{k}_2(\omega)$ relative to its static value $\tilde{k}_2(0) = k_2$.

\subsection{Structure of the paper} 

This completes the summary of our main results. In the remaining sections of the paper we present the calculational details behind these results. 

The spacetime of an unperturbed compact star is constructed in Sec.~\ref{sec:unpert}, and the tidal perturbation of the interior configuration is examined in Secs.~\ref{sec:pert_conf} and \ref{sec:pert_eqns}. The surface of the deformed star is identified in Sec.~\ref{sec:surface} and the junction conditions satisfied by the spacetime metric at this boundary are formulated in Sec.~\ref{sec:junction}; a part of this discussion is relegated to Appendix~\ref{sec:extrinsic}. The tidal constant $p_2$ is computed in Sec.~\ref{sec:love} by matching the interior metric constructed here to the exterior metric previously obtained in Ref.~\cite{poisson:21a}. In Sec.~\ref{sec:response}, the final section of the paper, we derive the results presented in Sec.~\ref{intro:response}. 

\section{Unperturbed configuration}
\label{sec:unpert}

Throughout this paper we examine a material body of mass $M$ and radius $R$ placed within a tidal environment created by one or more remote bodies. In this section we examine the unperturbed state of the star, when it is removed from the tidal environment. The star's tidal deformation will be considered next. 

The unperturbed star is a static and spherically symmetric configuration of perfect fluid, with a gravitational field described in general relativity. We write the interior metric as
\begin{equation}
ds^2 = -e^{2\psi}\, dt^2 + (1-2m/r)^{-1}\, dr^2 + r^2 \bigl( d\theta^2 + \sin^2\theta\, d\phi^2 \bigr),
\label{metric_unpert}
\end{equation}
with a potential $\psi$ and a mass function $m$ that depend on the radial coordinate $r$.

The state of the fluid is described by a rest-mass density $\rho$ (the product of the number density of the constituent particles times the rest mass of each particle), an internal energy density $\varepsilon$, a total energy density $\mu := \rho + \varepsilon$, a pressure $p$, and a velocity field $u^\alpha$. The fluid's energy-momentum tensor is given by
\begin{equation}
T^{\alpha\beta} = \mu\, u^\alpha u^\beta + p \bigl( g^{\alpha\beta} + u^\alpha u^\beta \bigr).
\end{equation}
We assume that the fluid comes with a zero-temperature equation of state of the form $\varepsilon = \varepsilon(\rho)$. The pressure is then defined via the first law of thermodynamics, $d(\varepsilon/\rho) + p\, d(1/\rho) = 0$, and it is also a function of $\rho$. It is useful to introduce the specific enthalpy $h$, defined by
\begin{equation}
h := \frac{\mu + p}{\rho} = 1 + \frac{\varepsilon + p}{\rho}. 
\label{h_def} 
\end{equation}
The first law can then be expressed as $d\mu = h\, d\rho$, and Eq.~(\ref{h_def}) implies that $h^{-1}\, dh = (\mu+p)^{-1}\, dp$.

For a static and spherically symmetric configuration, the fluid variables depend on $r$ only, and the only nonvanishing component of the velocity field is $u^t = e^{-\psi}$. The Einstein field equations produce the differential equations
\begin{equation}
\frac{d\psi}{dr} =\frac{m + 4\pi r^3 p}{r^2 (1-2m/r)}, \qquad
\frac{dm}{dr} = 4\pi r^2 \mu
\label{field_eqns}
\end{equation}
for the potential and mass function. The statement of hydrostatic equilibrium is
\begin{equation}
\frac{dp}{dr} = -(\mu + p) \frac{d\psi}{dr} = -\frac{(\mu+p)(m + 4\pi r^3 p)}{r^2 (1-2m/r)}.
\label{hydro_eq}
\end{equation}
Given a choice of equation of state, these equations are integrated to obtain a stellar model. Integration proceeds with a chosen value of the central density $\rho_c := \rho(r=0)$, until the pressure vanishes at the stellar surface, $p(r=R) = 0$. Outside the star the metric functions take on their Schwarzschild expressions, $e^{2\psi} = 1-2M/r$ and $m = M$, with $M := m(r=R)$ denoting the star's gravitational mass. 

\section{Perturbed configuration}
\label{sec:pert_conf}

We now place the star within a tidal environment, and aim to describe its deformation through second order in the tidal displacement. In our developments we may ignore the time dependence of the tidal field, and therefore idealize it as a static field. The dynamical aspects of the tidal interaction were considered in Ref.~\cite{pitre-poisson:24} and incorporated in the discussion of Sec.~\ref{intro:response}. 

At first order in the tidal interaction, the tidal field is dominated by its leading term involving the quadrupole moment ${\cal E}_{ab}$, a symmetric-tracefree Cartesian tensor. But we also include, for technical reasons that will become clear in Sec.~\ref{sec:love}, a subdominant term involving the hexadecapole moment ${\cal E}_{abcd}$, another symmetric-tracefree Cartesian tensor. When combined with $\Omega^a$, the unit radial vector of Eq.~(\ref{Omega_def}), the tidal moments become spherical harmonics of degree $\ell=2$ and $\ell=4$, respectively. To describe these we use the notation
\begin{equation}
\Eq := {\cal E}_{ab}\, \Omega^a \Omega^b, \qquad
\Eh := {\cal E}_{abcd}\, \Omega^a \Omega^b \Omega^c \Omega^d. 
\end{equation}
At second order the tidal field couples with itself, and we describe this coupling by introducing terms quadratic in ${\cal E}_{ab}$; we ignore higher-order couplings that would involve ${\cal E}_{abcd}$. The irreducible quadratic pieces are
\begin{equation}
{\cal E}_{ab} {\cal E}^{ab}, \qquad
{\cal E}_{c\langle a} {\cal E}^c_{\ b\rangle}, \qquad
{\cal E}_{\langle ab} {\cal E}_{cd \rangle},
\end{equation}
in which the angular brackets are an instruction to remove all traces. The corresponding spherical harmonics are 
\begin{equation}
\EEm := {\cal E}_{ab} {\cal E}^{ab}, \qquad
\EEq := {\cal E}_{c\langle a} {\cal E}^c_{\ b\rangle}\, \Omega^a \Omega^b, \qquad
\EEh := {\cal E}_{\langle ab} {\cal E}_{cd \rangle}\, \Omega^a \Omega^b \Omega^c \Omega^d, 
\end{equation} 
and they come with $\ell = 0$, $\ell = 2$, and $\ell = 4$, respectively. The labels on $\cal E$ and $({\cal EE})$ refer to the multipole order: monopole ({\sf m}) for $\ell = 0$, quadrupole ($\sf q$) for $\ell = 2$, and hexadecapole ($\sf h$) for $\ell = 4$. 

The tidal deformation forces all fluid and metric variables to acquire a perturbation with respect to the spherically symmetric state. We write the perturbed metric as $g_{\alpha\beta} + \delta g_{\alpha\beta}$, with $g_{\alpha\beta}$ denoting the unperturbed metric of Eq.~(\ref{metric_unpert}), while 
\begin{subequations}
\label{pert_metric}
\begin{align}
\delta g_{tt} &= 2 e^{2\psi} \biggl\{ \frac{R^3}{M} \ovr{U}{1}{q}\, \Eq
+ \frac{R^5}{M} \ovr{U}{1}{h}\, \Eh
+ \frac{R^6}{M^2} \Bigl[ \ovr{U}{2}{m}\, \EEm + \ovr{U}{2}{q}\, \EEq + \ovr{U}{2}{h}\, \EEh \Bigr] \biggr\},\\
\delta g_{rr} &= \frac{2}{1-2m/r} \biggl\{ \frac{R^3}{M} \ovr{V}{1}{q}\, \Eq
+ \frac{R^5}{M} \ovr{V}{1}{h}\, \Eh
+ \frac{R^6}{M^2} \Bigl[ \ovr{V}{2}{m}\, \EEm + \ovr{V}{2}{q}\, \EEq + \ovr{V}{2}{h}\, \EEh \Bigr] \biggr\}, \\
\delta g_{AB} &= 2 r^2 \Omega_{AB} \biggl\{ \frac{R^3}{M} \ovr{W}{1}{q}\, \Eq
+ \frac{R^5}{M} \ovr{W}{1}{h}\, \Eh
+ \frac{R^6}{M^2} \Bigl[ \ovr{W}{2}{q}\, \EEq + \ovr{W}{2}{h}\, \EEh \Bigr] \biggr\},
\end{align}
\end{subequations}
represents the perturbation, which includes both first-order and second-order terms. The various coefficients $\ovr{U}{j}{\ell}$, $\ovr{V}{j}{\ell}$, $\ovr{W}{j}{\ell}$ are functions of $r$ only --- in this notation, $j = \{1 , 2\}$ stands for the order of the perturbation and $\ell = \{ {\sf m}, {\sf q}, {\sf h} \}$ for its multipolar order --- and the factors of $R$ and $M$ are inserted to ensure that each radial function is dimensionless. We use an uppercase latin index $A = (2,3)$ to refer to the angular coordinates $\theta^A := (\theta,\phi)$, and $\Omega_{AB} := \mbox{diag}[1, \sin^2\theta]$ denotes the metric on a unit two-sphere.

For the $\ell = 2$ and $\ell = 4$ terms in Eq.~(\ref{pert_metric}), the metric perturbation satisfies the Regge-Wheeler gauge condition \cite{regge-wheeler:57, martel-poisson:05}, which sets $\delta g_{tA} = \delta g_{rA} = 0$ and eliminates a tracefree contribution to $\delta g_{AB}$. This gauge, however, is not defined for $\ell = 0$, and in this case we fix the gauge by setting to zero a term proportional to $\EEm$ in $\delta g_{AB}$. We find it convenient to write
\begin{equation}
\ovr{W}{j}{\ell} = \ovr{U}{j}{\ell} + \ovr{K}{j}{\ell}
\end{equation}
for some new variables $\ovr{K}{j}{\ell}$, and the Einstein field equations imply that $\ovr{V}{j}{\ell}$ is given by
\begin{subequations}
\begin{align} 
\ovr{V}{1}{q} &= \ovr{U}{1}{q}, \\
\ovr{V}{1}{h} &= \ovr{U}{1}{h}, \\
\ovr{V}{2}{q} &= \ovr{U}{2}{q} - \frac{4}{7}\bigl( \ovr{U}{1}{q} \bigr)^2, \\
\ovr{V}{2}{h} &= \ovr{U}{2}{h} + \frac{4}{3} \bigl( \ovr{U}{1}{q} \bigr)^2.
\end{align} 
\end{subequations}
The independent radial functions are $\ovr{U}{j}{\ell}$, $\ovr{K}{j}{\ell}$, and $\ovr{V}{2}{m}$.

The fluid's velocity field acquires a perturbation $\delta u^\alpha$, with nonvanishing component
\begin{equation}
\delta u^t = e^{-\psi} \biggl\{ \frac{R^3}{M} \ovr{u}{1}{q}\, \Eq
+ \frac{R^5}{M} \ovr{u}{1}{h}\, \Eh
+ \frac{R^6}{M^2} \Bigl[ \ovr{u}{2}{m}\, \EEm + \ovr{u}{2}{q}\, \EEq + \ovr{u}{2}{h}\, \EEh \Bigr]\biggr\}.
\end{equation}
Normalization of the perturbed velocity $u^\alpha + \delta u^\alpha$ in the perturbed metric implies that
\begin{subequations}
\begin{align} 
\ovr{u}{1}{q} &= \ovr{U}{1}{q}, \\
\ovr{u}{1}{h} &= \ovr{U}{1}{h}, \\
\ovr{u}{2}{m} &= \ovr{U}{2}{m} + \frac{1}{5} \bigl( \ovr{U}{1}{q} \bigr)^2, \\
\ovr{u}{2}{q} &= \ovr{U}{2}{q} + \frac{6}{7} \bigl( \ovr{U}{1}{q} \bigr)^2, \\
\ovr{u}{2}{h} &= \ovr{U}{2}{h} + \frac{3}{2} \bigl( \ovr{U}{1}{q} \bigr)^2.
\end{align} 
\end{subequations}
The perturbations in the energy density and pressure are decomposed as
\begin{equation}
\delta \mu = \frac{R^3}{M} \ovr{\mu}{1}{q}\, \Eq
+ \frac{R^5}{M} \ovr{\mu}{1}{h}\, \Eh
+ \frac{R^6}{M^2} \Bigl[ \ovr{\mu}{2}{m}\, \EEm + \ovr{\mu}{2}{q}\, \EEq + \ovr{\mu}{2}{h}\, \EEh \Bigr]
\label{del_mu} 
\end{equation}
and
\begin{equation}
\delta p = \frac{R^3}{M} \ovr{p}{1}{q}\, \Eq
+ \frac{R^5}{M} \ovr{p}{1}{h}\, \Eh
+ \frac{R^6}{M^2} \Bigl[ \ovr{p}{2}{m}\, \EEm + \ovr{p}{2}{q}\, \EEq + \ovr{p}{2}{h}\, \EEh \Bigr].
\label{del_p} 
\end{equation}
The coefficients $\ovr{\mu}{j}{\ell}$ are associated with $\ovr{p}{j}{\ell}$ by virtue of the equation of state, which we can write in the form of $\mu = \mu(p)$. This implies that 
\begin{equation}
\delta \mu = \frac{d\mu}{dp}\, \delta p + \frac{1}{2} \frac{d^2\mu}{dp^2}\, \delta p^2 + \cdots,
\end{equation}
and substitution of Eqs.~(\ref{del_mu}) and (\ref{del_p}) reveals that
\begin{subequations}
\begin{align} 
\ovr{\mu}{1}{q} &= \frac{d\mu}{dp}\, \ovr{p}{1}{q}, \\
\ovr{\mu}{1}{h} &= \frac{d\mu}{dp}\, \ovr{p}{1}{h}, \\
\ovr{\mu}{2}{m} &= \frac{d\mu}{dp}\, \ovr{p}{2}{m}
+ \frac{1}{15} \frac{d^2\mu}{dp^2} \bigl( \ovr{p}{1}{q} \bigr)^2, \\ 
\ovr{\mu}{2}{q} &= \frac{d\mu}{dp}\, \ovr{p}{2}{q}
+ \frac{2}{7} \frac{d^2\mu}{dp^2} \bigl( \ovr{p}{1}{q} \bigr)^2, \\
\ovr{\mu}{2}{h} &= \frac{d\mu}{dp}\, \ovr{p}{2}{h}
+ \frac{1}{2} \frac{d^2\mu}{dp^2} \bigl( \ovr{p}{1}{q} \bigr)^2.
\end{align} 
\end{subequations}
The conservation statement $\nabla_\beta T^{\alpha\beta} = 0$ for the perturbed energy-momentum tensor in the perturbed spacetime further implies that $\ovr{p}{j}{\ell}$ is directly related to $\ovr{U}{j}{\ell}$. We have that
\begin{subequations}
\label{delp_coeffs} 
\begin{align} 
\ovr{p}{1}{q} &= (\mu+p)\, \ovr{U}{1}{q}, \\
\ovr{p}{1}{h} &= (\mu+p)\, \ovr{U}{1}{h}, \\
\ovr{p}{2}{q} &= (\mu+p) \Biggl[ \ovr{U}{2}{q}
+ \frac{2}{7} \biggl( 3 + \frac{d\mu}{dp} \biggr) \bigl( \ovr{U}{1}{q} \bigr)^2 \Biggr], \\
\ovr{p}{2}{h} &= (\mu+p) \Biggl[ \ovr{U}{2}{h}
+ \frac{1}{2} \biggl( 3 + \frac{d\mu}{dp} \biggr) \bigl( \ovr{U}{1}{q} \bigr)^2 \Biggr].
\end{align} 
\end{subequations}
We observe, however, that $\ovr{p}{2}{m}$ remains as an independent variable. 

The perturbed state of the body is described by a fairly small number of independent variables. At the linear level we require $\ovr{U}{1}{q}$ and $\ovr{K}{1}{q}$, as well as $\ovr{U}{1}{h}$ and $\ovr{K}{1}{h}$. At the quadratic level we need $\ovr{U}{2}{m}$, $\ovr{V}{2}{m}$, and $\ovr{p}{2}{m}$ for the monopole sector, $\ovr{U}{2}{q}$ and $\ovr{K}{2}{q}$ for the quadrupole sector, and $\ovr{U}{2}{h}$ and $\ovr{K}{2}{h}$ for the hexadecapole sector. Henceforth we shall focus our attention on the $\ell = 2$ and $\ell = 4$ pieces of the perturbation and completely ignore the $\ell = 0$ part, which delivers no useful information.\footnote{As is shown in Sec.~V E of Ref.~\cite{poisson:21a}, the $\ell=0$ piece of the exterior metric contains two arbitrary constants, and each one can be set to zero without loss of generality (by a choice of gauge and a redefinition of the mass parameter, respectively). There is no tidal constant $p_0$ associated with the monopole sector of the nonlinear tidal perturbation.}

The manipulations presented in this section require the decomposition of $\E^{\sf q} \E^{\sf q} = \E_{ab} \E_{cd}\, \Omega^a \Omega^b \Omega^c \Omega^d$ into the irreducible pieces $\EEm$, $\EEq$, and $\EEh$. This is accomplished with   
\begin{equation} 
\E^{\sf q} \E^{\sf q} = \frac{2}{15} \EEm + \frac{4}{7} \EEq + \EEh, 
\label{EqEq} 
\end{equation} 
an identity established in Sec.~II B of Ref.~\cite{poisson:21a}. 

\section{Perturbation equations}
\label{sec:pert_eqns} 

The $rr$ and $rA$ components of the Einstein field equations for the perturbed configuration supply systems of first-order differential equations for the perturbation variables. The remaining components provide only redundant information.

At the linear level we obtain
\begin{subequations}
\label{lin_L2}
\begin{align} 
m_+ \frac{d}{dr}\, \ovr{U}{1}{q} &= \Biggl\{ 4\pi r^2 \mu_+
+ \frac{2}{f} \biggl[ \frac{m_+^2}{r^2} - \frac{(1+16\pi r^2 p) m_+}{r}
+ 4\pi r^2 p (1+8\pi r^2 p) \biggr] \Biggr\}\, \ovr{U}{1}{q} + 2\, \ovr{K}{1}{q}, \\
r \frac{d}{dr}\, \ovr{K}{1}{q} &= \frac{2m_+}{r f}\, \ovr{U}{1}{q}
\end{align}
\end{subequations}
for the quadropole sector, and
\begin{subequations}
\label{lin_L4}
\begin{align} 
m_+ \frac{d}{dr}\, \ovr{U}{1}{h} &= \Biggl\{ 4\pi r^2 \mu_+
+ \frac{2}{f} \biggl[ \frac{m_+^2}{r^2} - \frac{(1+16\pi r^2 p) m_+}{r}
+ 4\pi r^2 p (1+8\pi r^2 p) \biggr] \Biggr\}\, \ovr{U}{1}{h} + 9\, \ovr{K}{1}{h}, \\
r \frac{d}{dr}\, \ovr{K}{1}{h} &= \frac{2m_+}{r f}\, \ovr{U}{1}{h}
\end{align}
\end{subequations}
for the hexadecapole sector. At the quadratic level we have
\begin{subequations}
\label{quad_L2}
\begin{align} 
m_+ \frac{d}{dr}\, \ovr{U}{2}{q} &= \Biggl\{ 4\pi r^2 \mu_+
+ \frac{2}{f} \biggl[ \frac{m_+}{r^2} - \frac{(1+16\pi r^2 p) m_+}{r}
+ 4\pi r^2 p (1+8\pi r^2 p) \biggr] \Biggr\}\, \ovr{U}{2}{q} + 2\, \ovr{K}{2}{q}
\nonumber \\ & \quad \mbox{} 
+ \Biggl\{ \frac{32\pi^2 r^6 f}{7 m_+^2}\, \mu_+^2
+ \frac{8}{7} \biggl[ \pi r^2 \Bigl(5 + \frac{d\mu}{dp} \Bigr)
- \frac{2\pi r^3(1 + 24\pi r^2 p)}{m_+}  
+ \frac{16\pi^2 r^6 p(1 + 8\pi r^2 p)}{m_+^2} \biggr]\, \mu_+
\nonumber \\ & \quad \mbox{} 
- \frac{2}{7 f} \biggl[ \frac{2m_+^2}{r^2} - \frac{26m_+}{r} - 640 \pi^2 r^4 p^2
+ 40 \pi r^2 p + 11 + \frac{16\pi r^3 p(1+8\pi r^2 p)(1 + 24\pi r^2 p)}{m_+}
\nonumber \\ & \quad \mbox{} 
- \frac{64\pi^2 r^6 p^2(1 + 8\pi r^2 p)^2}{m_+^2} \biggr]
\Biggr\}\, \bigl( \ovr{U}{1}{q} \bigr)^2
\nonumber \\ & \quad \mbox{} 
+ \frac{8}{7} \biggl[ \frac{4\pi r^4 f}{m_+^2}\, \mu_+
- 3 - \frac{r(1+24\pi r^2 p)}{m_+}
+ \frac{8\pi r^4 p(1 + 8\pi r^2 p)}{m_+^2} \biggr]\, \bigl( \ovr{U}{1}{q}\, \ovr{K}{1}{q} \bigr) 
\nonumber \\ & \quad \mbox{} 
- \frac{4}{7} \biggl[ 7 + \frac{4 r}{m_+} - \frac{2r^2(1+8\pi r^2 p)}{m_+^2}
\biggr]\, \bigl( \ovr{K}{1}{q} \bigr)^2, \\
r \frac{d}{dr}\, \ovr{K}{2}{q} &= \frac{2m_+}{r f}\, \ovr{U}{2}{q}
+ \frac{4}{7 f} \biggl[ \frac{12\pi r^3 f}{m_+}\, \mu_+
+ \frac{15 m_+}{r} - 3(3+40\pi r^2 p) + \frac{24\pi r^3 p(1+8\pi r^2 p)}{m_+}
\biggr]\, \bigl( \ovr{U}{1}{q} \bigr)^2
\nonumber \\ & \quad \mbox{} 
+ \frac{8}{7} \biggl[ \frac{4 \pi r^3}{m_+}\, \mu_+
- 2 + \frac{r(3+ 8\pi r^2 p)}{m_+} \biggr]\, \bigl( \ovr{U}{1}{q}\, \ovr{K}{1}{q} \bigr)
+ \frac{16 r}{7 m_+}\, \bigl( \ovr{K}{1}{q} \bigr)^2 
\end{align}
\end{subequations}
for the quadrupole sector, and 
\begin{subequations}
\label{quad_L4}
\begin{align} 
m_+ \frac{d}{dr}\, \ovr{U}{2}{h} &= \Biggl\{ 4\pi r^2 \mu_+
+ \frac{2}{f} \biggl[ \frac{m_+}{r^2} - \frac{(1+16\pi r^2 p) m_+}{r}
+ 4\pi r^2 p (1+8\pi r^2 p) \biggr] \Biggr\}\, \ovr{U}{2}{h} + 9\, \ovr{K}{2}{h}
\nonumber \\ & \quad \mbox{} 
+ \Biggl\{ \frac{8\pi^2 r^6 f}{m_+^2}\, \mu_+^2
+ 2 \biggl[ \pi r^2 \Bigl(5 + \frac{d\mu}{dp} \Bigr)
- \frac{2\pi r^3(1 + 24\pi r^2 p)}{m_+}  
+ \frac{16\pi^2 r^6 p(1 + 8\pi r^2 p)}{m_+^2} \biggr]\, \mu_+
\nonumber \\ & \quad \mbox{} 
- \frac{2}{f} \biggl[ \frac{5m_+^2}{3r^2} - \frac{17m_+}{r} - 160 \pi^2 r^4 p^2
+ 52 \pi r^2 p + 8
+ \frac{4\pi r^3 p(1+8\pi r^2 p)(1 + 24\pi r^2 p)}{m_+}
\nonumber \\ & \quad \mbox{} 
- \frac{16\pi^2 r^6 p^2(1 + 8\pi r^2 p)^2}{m_+^2} \biggr]
\Biggr\}\, \bigl( \ovr{U}{1}{q} \bigr)^2
\nonumber \\ & \quad \mbox{} 
+ 2\biggl[ \frac{4\pi r^4 f}{m_+^2}\, \mu_+
- 10 - \frac{r(1+24\pi r^2 p)}{m_+}
+ \frac{8\pi r^4 p(1 + 8\pi r^2 p)}{m_+^2} \biggr]\, \bigl( \ovr{U}{1}{q}\, \ovr{K}{1}{q} \bigr) 
\nonumber \\ & \quad \mbox{} 
- 2\biggl[ 7 + \frac{2 r}{m_+} - \frac{r^2(1+8\pi r^2 p)}{m_+^2}
\biggr]\, \bigl( \ovr{K}{1}{q} \bigr)^2, \\
r \frac{d}{dr}\, \ovr{K}{2}{h} &= \frac{2m_+}{r f}\, \ovr{U}{2}{h}
+ \frac{2}{3 f} \biggl[ \frac{18\pi r^3 f}{m_+}\, \mu_+
+ \frac{19 m_+}{r} - 2(5+76\pi r^2 p) + \frac{36\pi r^3 p(1+8\pi r^2 p)}{m_+}
\biggr]\, \bigl( \ovr{U}{1}{q} \bigr)^2
\nonumber \\ & \quad \mbox{} 
+ 2 \biggl[ \frac{4 \pi r^3}{m_+}\, \mu_+
- 2 + \frac{r(3+ 8\pi r^2 p)}{m_+} \biggr]\, \bigl( \ovr{U}{1}{q}\, \ovr{K}{1}{q} \bigr)
+ \frac{4 r}{m_+}\, \bigl( \ovr{K}{1}{q} \bigr)^2 
\end{align}
\end{subequations}
for the hexadecapole sector. To shorten the equations we introduced the notations $\mu_+ := \mu + p$, $m_+ := m + 4\pi r^3 p$, and $f := 1-2m/r$. We observe that the equations for the linear variables are homogeneous, while those for the quadratic variables feature $\ovr{U}{1}{q}$ and $\ovr{K}{1}{q}$ as source terms.

The differential equations for $\ovr{U}{2}{q}$ and $\ovr{U}{2}{h}$ feature a term proportional to $d\mu/dp$, which may become singular at $r=R$ for some equations of state --- we relegate a discussion of this eventuality to Appendix~\ref{sec:extrinsic}. To eliminate such terms, and to facilitate the numerical integration of the equations, we perform the change of variables
\begin{subequations}
\label{change} 
\begin{align}
\ovr{U}{1}{q} &= \frac{m + 4\pi r^3 p}{r}\, \ovr{a}{1}{q}, \\
\ovr{K}{1}{q} &=\frac{(m + 4\pi r^3 p)^2}{r^2}\, \ovr{b}{1}{q}, \\
\ovr{U}{1}{h} &= \frac{r(m + 4\pi r^3 p)}{R^2}\, \ovr{a}{1}{h}, \\
\ovr{K}{1}{h} &= \frac{(m + 4\pi r^3 p)^2}{R^2}\, \ovr{b}{1}{h}, \\
\ovr{U}{2}{q} &= \frac{m + 4\pi r^3 p}{r}\, \ovr{a}{2}{q}
- \frac{8\pi}{7} r^2 (1-2m/r) (\mu+p)\, \bigl( \ovr{a}{1}{q} \bigr)^2, \\  
\ovr{K}{2}{q} &= \frac{(m + 4\pi r^3 p)^2}{r^2}\, \ovr{b}{2}{q}, \\
\ovr{U}{2}{h} &= \frac{m + 4\pi r^3 p}{r}\, \ovr{a}{2}{h}
- 2\pi r^2 (1-2m/r) (\mu+p)\, \bigl( \ovr{a}{1}{q} \bigr)^2, \\  
\ovr{K}{2}{h} &= \frac{(m + 4\pi r^3 p)^2}{r^2}\, \ovr{b}{2}{h},
\end{align}
\end{subequations}
and utilize the new variables $\ovr{a}{j}{\ell}$ and $\ovr{b}{j}{\ell}$ in the numerical work. We shall not display the explicit form of the differential equations for these variables. 

An analysis of the differential equations near $r = 0$ reveals that the original variables behave as
\begin{equation}
\ovr{U}{1}{q} \sim r^2, \qquad
\ovr{U}{1}{h} \sim r^4, \qquad
\ovr{U}{2}{q} \sim r^2, \qquad
\ovr{U}{2}{h} \sim r^4
\end{equation}
and
\begin{equation}
\ovr{K}{1}{q} \sim r^4, \qquad
\ovr{K}{1}{h} \sim r^6, \qquad
\ovr{K}{2}{q} \sim r^4, \qquad
\ovr{K}{2}{h} \sim r^4
\end{equation}
near the stellar center. On the other hand, the new variables all approach a nonzero constant as $r \to 0$ and are therefore well suited to the numerical integrations. 

\section{Deformed surface}
\label{sec:surface}

The interior perturbation constructed in the preceding sections must be joined with the exterior perturbation described in Ref.~\cite{poisson:21a}. The transition occurs at the surface of the deformed star, and our task in this section is to locate this surface. The junction conditions at the stellar boundary will be considered next in Sec.~\ref{sec:junction}. 

The surface of the tidally deformed star is described by $r = R + \delta R$, with a deformation decomposed as
\begin{equation}
\delta R = \frac{R^4}{M}\, \ovr{r}{1}{q}\, \Eq + \frac{R^6}{M}\, \ovr{r}{1}{h}\, \Eh
+ \frac{R^7}{M^2} \Bigl[ \ovr{r}{2}{m}\, \EEm + \ovr{r}{2}{q}\, \EEq + \ovr{r}{2}{h}\, \EEh \Bigr],
\label{del_R} 
\end{equation}
where the coefficients $\ovr{r}{j}{\ell}$ are constants. To determine where the surface is situated we appeal to the specific enthalpy of Eq.~(\ref{h_def}), which is necessarily equal to 1 at the surface. We express this as
\begin{equation} 
\bigl( \ln h + \delta \ln h \bigr) \Bigr|_{r = R + \delta R} = 0,
\label{surface}
\end{equation}
where $\delta \ln h$ is the perturbation in the logarithm of the specific enthalpy, which we decompose as
\begin{equation}
\delta \ln h = \frac{R^3}{M}\, \ovr{h}{1}{q}\, \Eq + \frac{R^5}{M}\, \ovr{h}{1}{h}\, \Eh
+ \frac{R^6}{M^2} \Bigl[ \ovr{h}{2}{m}\, \EEm + \ovr{h}{2}{q}\, \EEq + \ovr{h}{2}{h}\, \EEh \Bigr],
\label{del_lnh} 
\end{equation}
with coefficients $\ovr{h}{j}{\ell}$ that depend on $r$.

The fluid's equation of state can be expressed as $\ln h = (\ln h)(p)$, and $\delta \ln h$ can be related to the $\delta p$ of Eq.~(\ref{del_p}) by the equation
\begin{equation}
\delta \ln h = \frac{d\ln h}{dp}\, \delta p + \frac{1}{2} \frac{d^2\ln h}{dp^2}\, \delta p^2.
\end{equation}
Derivatives of $\ln h$ are evaluated with the help of the first law of thermodynamics, $d\ln h = (\mu+p)^{-1}\, dp$ --- refer to the discussion below Eq.~(\ref{h_def}). After also making use of Eq.~(\ref{delp_coeffs}), we find that
\begin{subequations}
\begin{align}
\ovr{h}{1}{q} &= \ovr{U}{1}{q}, \\
\ovr{h}{1}{h} &= \ovr{U}{1}{h}, \\
\ovr{h}{2}{m} &= (\mu+p)^{-1}\, \ovr{p}{2}{m}
- \frac{1}{15} \biggl( 1 + \frac{d\mu}{dp} \biggr) \bigl( \ovr{U}{1}{q} \bigr)^2, \\
\ovr{h}{2}{q} &= \ovr{U}{2}{q} + \frac{4}{7} \bigl( \ovr{U}{1}{q} \bigr)^2, \\
\ovr{h}{2}{h} &= \ovr{U}{2}{h} + \bigl( \ovr{U}{1}{q} \bigr)^2.
\end{align}
\end{subequations}

We now return to Eq.~(\ref{surface}), which we expand through the second order in the deformation. The equation becomes
\begin{equation}
\frac{d\ln h}{dr} \biggr|_{r=R}\, \delta R
+ \frac{1}{2} \frac{d^2 \ln h}{dr^2} \biggr|_{r = R}\, \delta R^2
+ \delta \ln h \biggr|_{r=R} + \frac{\partial \delta \ln h}{\partial r} \biggr|_{r=R}\, \delta R
= 0.
\end{equation}
We then substitute Eqs.~(\ref{del_R}) and (\ref{del_lnh}), evaluate the radial derivatives of $\ln h$ with the help of Eq.~(\ref{hydro_eq}), and solve for the constants $\ovr{r}{j}{\ell}$. We eventually obtain 
\begin{subequations}
\begin{align}
\ovr{r}{1}{q} &= \frac{R}{M} \biggl( 1 - \frac{2M}{R} \biggr)\, \ovr{U}{1}{q}, \\
\ovr{r}{1}{h} &= \frac{R}{M} \biggl( 1 - \frac{2M}{R} \biggr)\, \ovr{U}{1}{h}, \\
\ovr{r}{2}{m} &= \frac{R}{M} \biggl( 1 - \frac{2M}{R} \biggr) \biggl[ (\mu+p)^{-1}\, \ovr{p}{2}{m}
+ \frac{1}{15} \biggl( 1 - \frac{2M}{R} - \frac{d\mu}{dp} \biggr) \bigl( \ovr{U}{1}{q} \bigr)^2
+ \frac{4}{15} \biggl( \frac{R}{M} \biggr)^2 \biggl( 1 - \frac{2M}{R} \biggr)\,
\bigl( \ovr{U}{1}{q}\, \ovr{K}{1}{q} \bigr) \biggr], \\
\ovr{r}{2}{q} &= \frac{R}{M} \biggl( 1 - \frac{2M}{R} \biggr) \biggl[ \ovr{U}{2}{q} 
- \frac{4}{7} \frac{R}{M} \biggl( 1 - \frac{2M}{R} \biggr) \bigl( \ovr{U}{1}{q} \bigr)^2
+ \frac{8}{7} \biggl( \frac{R}{M} \biggr)^2 \biggl( 1 - \frac{2M}{R} \biggr)\,
\bigl( \ovr{U}{1}{q}\, \ovr{K}{1}{q} \bigr) \biggr], \\
\ovr{r}{2}{h} &= \frac{R}{M} \biggl( 1 - \frac{2M}{R} \biggr) \biggl[ \ovr{U}{2}{h} 
- \frac{R}{M} \biggl( 1 - \frac{2M}{R} \biggr) \bigl( \ovr{U}{1}{q} \bigr)^2
+ 2 \biggl( \frac{R}{M} \biggr)^2 \biggl( 1 - \frac{2M}{R} \biggr)\,
\bigl( \ovr{U}{1}{q}\, \ovr{K}{1}{q} \bigr) \biggr],
\end{align}
\end{subequations}
in which the perturbation variables are all evaluated at $r = R$. In the case of $\ovr{r}{2}{m}$ the right-hand side of the equation must be interpreted with care, as it involves quantities such as $(\mu + p)^{-1}$ and $d\mu/dp$, which may be singular at $r=R$ for some equations of state --- we again postpone a discussion of this point to Appendix~\ref{sec:extrinsic}. We shall not be preoccupied with this issue (which is likely to be resolved with a change of perturbation variables, analogous to what was done in Sec.~\ref{sec:pert_eqns}), as we are not interested in the monopole sector of the quadratic perturbation.

\section{Junction conditions}
\label{sec:junction}

The interior metric of a tidally deformed body, constructed in Secs.~\ref{sec:pert_conf} and \ref{sec:pert_eqns} in terms of the perturbation variables $\ovr{U}{j}{\ell}$ and $\ovr{K}{j}{\ell}$, must connect smoothly with the exterior metric obtained in Ref.~\cite{poisson:21a}; the joint occurs at the deformed surface described in Sec.~\ref{sec:surface}. The proper geometric formulation of the junction conditions was provided by Israel \cite{israel:66}, and we demand that the intrinsic metric and extrinsic curvature of the boundary surface be the same when evaluated on the interior and exterior sides. In symbols, we insist that
\begin{equation}
\bigl[ h_{ab} \bigr] = 0 = \bigl[ K_{ab} \bigr],
\label{israel} 
\end{equation}
where $h_{ab}$ is the induced metric, $K_{ab}$ is the extrinsic curvature, and $[ \psi ]$ denotes the jump of a quantity $\psi$ across the boundary surface (its value evaluated on the exterior side minus its value on the interior side).

When viewed from the stellar interior, the boundary surface is described by the parametric equations $x^\alpha = x^\alpha(y^a)$, in which $x^\alpha$ are the coordinates used in the interior spacetime and $y^a$ are intrinsic coordinates on the surface. We adopt $y^a = (t,\theta,\phi)$, and the explicit listing of equations is $t = t$, $r = R + \delta R$, $\theta = \theta$, and $\phi = \phi$. The vectors $e^\alpha_a := \partial x^\alpha/\partial y^a$ are tangent to the surface, and $n_\alpha \propto \partial_\alpha (r - R - \delta R)$ is the unit normal vector. The induced metric and extrinsic curvature are defined by
\begin{equation}
h_{ab} := e^\alpha_a e^\beta_b \bigl( g_{\alpha\beta} + \delta g_{\alpha\beta} \bigr), \qquad
K_{ab} := e^\alpha_a e^\beta_b \nabla_\alpha n_\beta,
\label{h_K}
\end{equation}
in which $\nabla_\alpha$ is the covariant-derivative operator compatible with the perturbed metric. It is understood that in Eq.~(\ref{h_K}), the right-hand sides are evaluated on the boundary surface and expressed in terms of the intrinsic coordinates.

The description is exactly the same when the boundary surface is viewed from the stellar exterior, except that the perturbed metric is now the one obtained in Ref.~\cite{poisson:21a}. For our purposes here we express $\delta g_{\alpha\beta}$ as in Eq.~(\ref{pert_metric}), except that we replace the prefactor $e^{2\psi}$ in $\delta g_{tt}$ by its Schwarzschild expression $1-2M/r$, and the factor $(1-2m/r)^{-1}$ in front of $\delta g_{rr}$ by $(1-2M/r)^{-1}$; here $M := m(r=R)$ is the body's gravitational mass. The radial functions $\ovr{U}{j}{\ell}$ and $\ovr{K}{j}{\ell}$ that occur in the exterior metric were obtained as explicit analytical expressions in Ref.~\cite{poisson:21a}.

We shall not present the detail of these computations, but merely state the conclusion that the junction conditions $[h_{ab}] = 0$ enforce the equality of the interior and exterior radial functions at $r=R$:
\begin{equation}
\ovr{U}{j}{\ell}^{\rm int}(r=R) = \ovr{U}{j}{\ell}^{\rm ext}(r=R), \qquad
\ovr{K}{j}{\ell}^{\rm int}(r=R) = \ovr{K}{j}{\ell}^{\rm ext}(r=R).
\label{junction} 
\end{equation}
In view of the first-order character of the differential equations listed in Sec.~\ref{sec:pert_eqns}, these are the only conditions that are required to identify a unique solution to the global problem, which consists of finding interior and exterior solutions to the perturbation equations and joining them smoothly at the boundary surface.

It is interesting to also investigate the junction conditions associated with the extrinsic curvature. We carry this out in Appendix~\ref{sec:extrinsic}. 

\section{Determination of the tidal constants}
\label{sec:love} 

The numerical integration of the structure equations listed in Sec.~\ref{sec:unpert} and the perturbation equations displayed in Sec.~\ref{sec:pert_eqns} requires a specific choice of equation of state for the matter that makes up the star. We adopt a polytropic model for which the pressure and internal energy density are respectively given by 
\begin{equation}
p = K \rho^{1+1/n}, \qquad \varepsilon = n p,
\label{polytrope} 
\end{equation}
where $\rho$ is the rest-mass density, while $K$ and $n$ are constants. It is helpful to introduce new, dimensionless variables to stand for the density $\rho$, mass function $m$, and radial coordinate $r$; this is described in detail in Sec.~IV D of Ref.~\cite{pitre-poisson:24}, and we shall not repeat this discussion here. It suffices to say that given a choice of polytropic index $n$, a numerical integration of the structure equations (\ref{field_eqns}) and (\ref{hydro_eq}) produces a stellar model parametrized by $\rho_c := \rho(r=0)$, the central value of the mass density; varying $\rho_c$ creates a sequence of equilibrium configurations. For each member of the sequence, the surface at $r=R$ is where $\rho$ (and therefore $p$) first vanishes, and $M := m(r=R)$ is the star's gravitational mass. The sequence terminates at the configuration of maximum mass; beyond this point the stellar models are dynamically unstable to radial perturbations.

The numerical integration of the perturbation equations, written in terms of the variables $\ovr{a}{j}{\ell}$ and $\ovr{b}{j}{\ell}$, proceeds from the stellar center at $r=0$. The equations, however, are singular at $r = 0$, and a Taylor expansion of all variables in powers of $r^2$ is exploited to provide starting values at a small but nonvanishing value of $r$. Because each variable approaches a constant as $r \to 0$, we write
\begin{equation}
\ovr{a}{j}{\ell} = \ovr{a}{j}{\ell}^0 + \ovr{a}{j}{\ell}^2\, r^2 + \ovr{a}{j}{\ell}^4\, r^4 + \cdots, \qquad
\ovr{b}{j}{\ell} = \ovr{b}{j}{\ell}^0 + \ovr{b}{j}{\ell}^2\, r^2 + \ovr{b}{j}{\ell}^4\, r^4 + \cdots
\label{expansions_r=0} 
\end{equation}
and use the differential equations to find relationships among the constant coefficients; the superscript on $\ovr{a}{j}{\ell}$ and $\ovr{b}{j}{\ell}$ indicates the power of $r$ to which each coefficient is attached. The integration continues to the stellar surface at $r=R$, and the junction conditions of Eq.~(\ref{junction}) relate the interior perturbation variables to their exterior analogues; these are defined in Secs.~V C and V E of Ref.~\cite{poisson:21a} and listed explicitly in its Appendix~C and D.

We first examine the quadrupole sector of the linear perturbation, governed by Eq.~(\ref{lin_L2}) re-expressed in terms of $\ovr{a}{1}{q}$ and $\ovr{b}{1}{q}$, which are defined by Eq.~(\ref{change}). An analysis of the equations near $r=0$ reveals that the coefficient $\ovr{a}{1}{q}^0$ in Eq.~(\ref{expansions_r=0}) is arbitrary, and that all other coefficients are determined in terms of this one. Because the differential equations are homogeneous, we have that $\ovr{a}{1}{q}^0$ is an overall multiplicative constant for the solution, and we can write
\begin{equation}
\ovr{a}{1}{q} = \ovr{a}{1}{q}^0\, \ovr{\hat{a}}{1}{q}, \qquad
\ovr{b}{1}{q} = \ovr{a}{1}{q}^0\, \ovr{\hat{b}}{1}{q},
\end{equation}
in which the hatted variables are solutions constructed by setting $\ovr{a}{1}{q}^0 = 1$. The exterior variables depend on another unknown constant, the tidal constant $k_2$, and the junction conditions supply two equations for the two unknowns. Integration of Eq.~(\ref{lin_L2}) and imposition of Eq.~(\ref{junction}) therefore allows us to determine both $\ovr{a}{1}{q}^0$ and $k_2$.

Identical considerations apply to Eq.~(\ref{lin_L4}), which governs the hexadecapole sector of the linear perturbation. Integration of these equations and imposition of the junction conditions delivers $\ovr{a}{1}{h}^0$ and $k_4$.

Next we deal with the quadrupole sector of the quadratic perturbation, governed by Eq.~(\ref{quad_L2}) rewritten in terms of $\ovr{a}{2}{q}$ and $\ovr{b}{2}{q}$, defined by Eq.~(\ref{change}). In this case we find that $\ovr{a}{2}{q}^0$ remains arbitrary in an analysis of the equations near $r=0$, and that it determines all the other coefficients. The freedom to adjust $\ovr{a}{2}{q}^0$ corresponds to the freedom of adding a solution to the homogeneous version of Eq.~(\ref{quad_L2}) --- obtained by setting $\ovr{a}{1}{q} = \ovr{b}{1}{q} = 0$ --- to any particular solution to the equations. On the other hand, the exterior solution, displayed in Eqs.~(5.25) and (5.26) of Ref.~\cite{poisson:21a}, depends on two unknown constants,\footnote{We recall the change of sign convention: the $p_2$ defined here is minus the $p_2$ defined in Ref.~\cite{poisson:21a}.} $T_2$ and $P_2 := -2 (R/M)^8\, p_2$. The constant $T_2$ is attached to a solution to the homogeneous version of the exterior equations, and the freedom to adjust it corresponds to the freedom to choose $\ovr{a}{2}{q}^0$ in the interior solution. Moreover, as discussed in detail in Sec.~V K of Ref.~\cite{poisson:21a}, $T_2$ can be changed at will by redefining the tidal moments ${\cal E}_{ab}$ according to 
\begin{equation}
{\cal E}_{ab}^{\rm old} = {\cal E}_{ab}^{\rm new}
+ \lambda_2 M^2\, {\cal E}^{\rm new}_{c\langle a} {\cal E}^{\rm{new}\, c}_{b\rangle},
\end{equation}
in which $\lambda_2$ is an arbitrary constant. The redefinition changes $T_2$ and $P_2$ according to
\begin{equation}
T_2^{\rm new} = T_2^{\rm old} + \lambda_2, \qquad
P_2^{\rm new} = P_2^{\rm old} + 2 \lambda_2 K_2,
\end{equation}
where $K_2 := (R/M)^5\, k_2$. We handle all this arbitrariness in the following way. We choose $\ovr{a}{2}{q}^0$ freely, construct the interior solution, and use the junction conditions of Eq.~(\ref{junction}) to determine $T_2$ and $P_2$. Our exterior solution, however, is required to be compatible with the metric of Eq.~(\ref{metric_sample}), which imposes the choice $T_2 = 0$. We therefore implement a redefinition of ${\cal E}_{ab}$, choose $\lambda_2$ so that $T^{\rm new}_2 = 0$, and calculate the new value of $P_2$. At this stage the freedom to redefine ${\cal E}_{ab}$ is exhausted, and the final value of $P_2$ is automatically invariant under further redefinitions.

\begin{figure}
\includegraphics[width=0.6\linewidth]{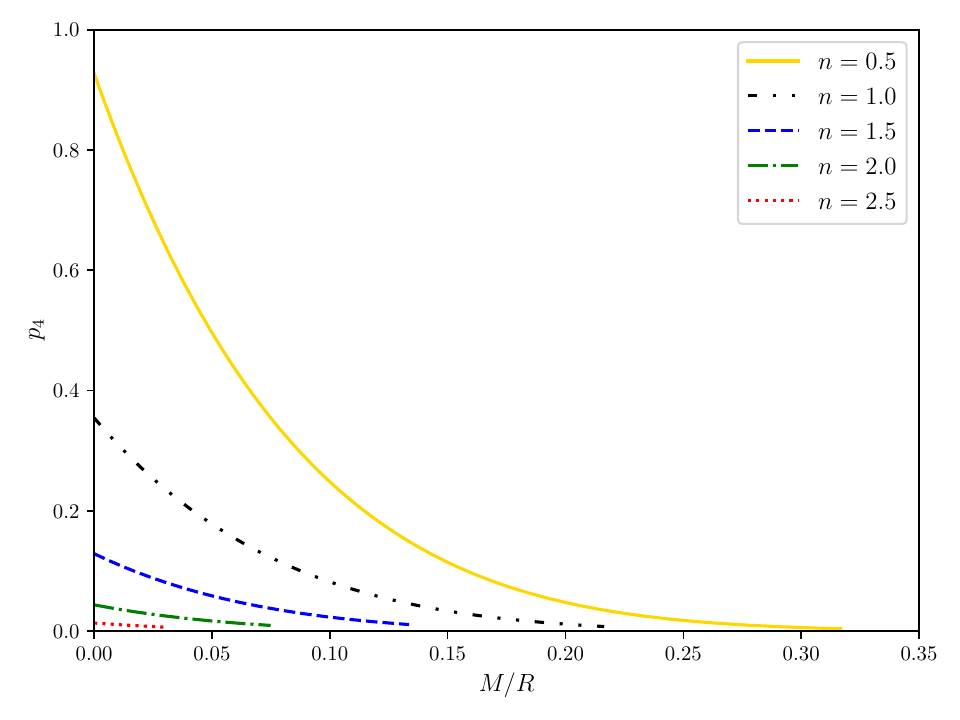}
\caption{Quadratic tidal constant $p_4$ as a function of stellar compactness $M/R$. The computation is carried out for polytropic stellar models with equation of state $p = K \rho^{1+1/n}$. Yellow curve: $n=0.5$. Black curve: $n=1$. Blue curve: $n=1.5$. Green curve: $n=2.0$. Red curve: $n=2.5$.}  
\label{fig:p4} 
\end{figure} 

\begin{figure}
\includegraphics[width=0.49\linewidth]{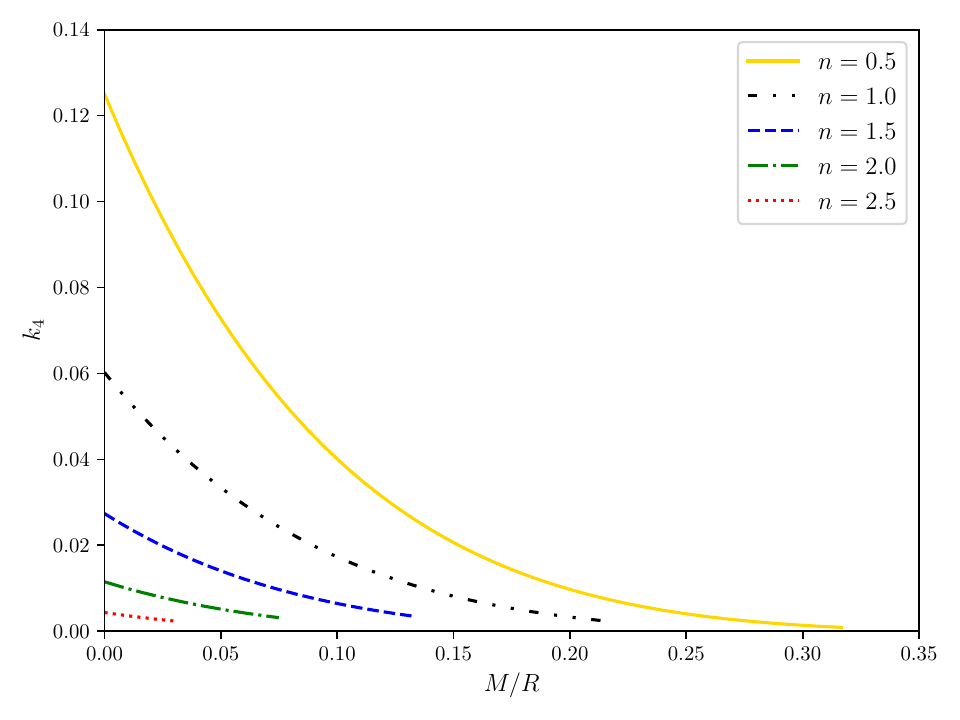}
\includegraphics[width=0.49\linewidth]{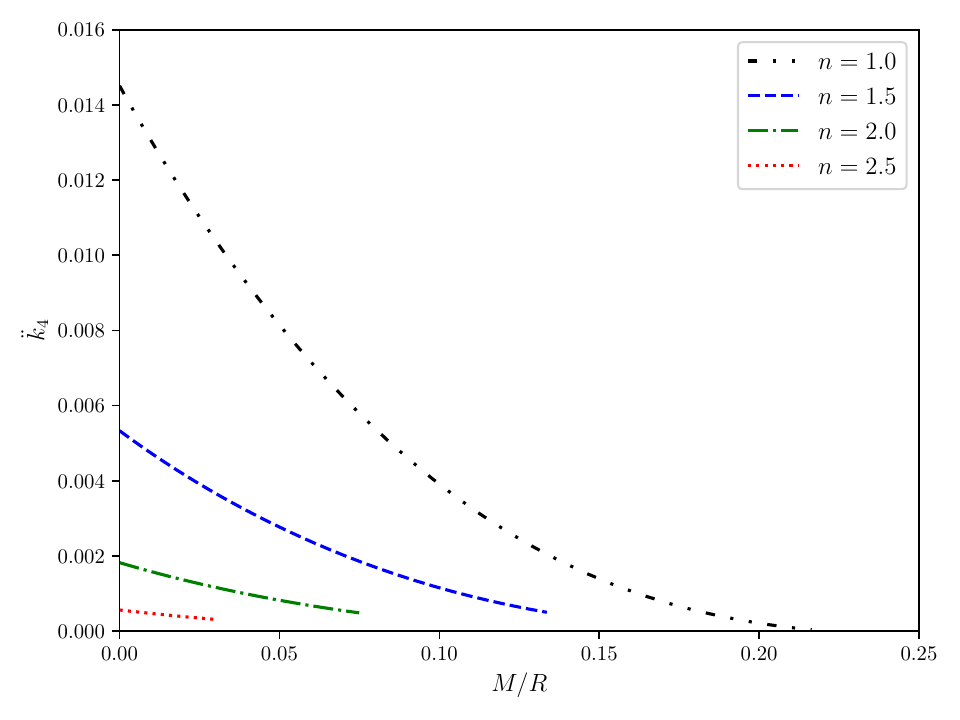}
\caption{Static and dynamic tidal constants as functions of stellar compactness $M/R$, computed for polytropes with equation of state $p = K \rho^{1+1/n}$. Left: Static constant $k_4$. Right: Dynamic constant $\ddot{k}_4$ (this figure is adapted from Ref.~\cite{pitre-poisson:24}). Yellow curve: $n=0.5$. Black curves: $n=1$. Blue curves: $n=1.5$. Green curves: $n=2.0$. Red curves: $n=2.5$.}  
\label{fig:k4} 
\end{figure} 

The hexadecapole sector of the quadratic perturbation is handled in an almost identical way. The perturbation equations are given by Eq.~(\ref{quad_L4}), re-expressed in terms of $\ovr{a}{2}{h}$ and $\ovr{b}{2}{h}$ defined by Eq.~(\ref{change}). It is now the coefficient $\ovr{a}{2}{h}^2$ that remains arbitrary in an analysis of the equations near $r=0$, and the freedom to adjust it corresponds to the freedom of adding a solution to the homogeneous version of Eq.~(\ref{quad_L4}). The exterior solution, given by Eqs.~(5.28) and (5.29) of Ref.~\cite{poisson:21a}, depends on two arbitrary constants, $T_4$ and $P_4 := -2 (R/M)^{10}\, p_4$, with $T_4$ attached to a solution to the homogeneous version of the exterior equations. This constant can be freely altered by redefining ${\cal E}_{abcd}$ according to 
\begin{equation}
{\cal E}_{abcd}^{\rm old} = {\cal E}_{abcd}^{\rm new}
+ \lambda_4\, {\cal E}^{\rm new}_{\langle ab} {\cal E}^{\rm new}_{bc\rangle}, 
\end{equation}
with $\lambda_4$ any constant. This changes $T_4$ and $P_4$ according to
\begin{equation}
T_4^{\rm new} = T_4^{\rm old} + \frac{1}{6} \lambda_4, \qquad
P_4^{\rm new} = P_4^{\rm old} + \frac{1}{3} \lambda_4 K_4,
\end{equation}
with $K_4 := (R/M)^9\, k_4$. We therefore pick any value for $\ovr{a}{2}{h}^2$ to construct the interior solution, and use Eq.~(\ref{junction}) to determine $T_4$ and $P_4$. To enforce $T_4 = 0$, as required by Eq.~(\ref{metric_sample}), we implement the redefinition of ${\cal E}_{abcd}$ and choose $\lambda_4$ so that $T^{\rm new}_4 = 0$. The new and final value of $P_4$ is then automatically invariant under further redefinitions of the tidal moments.   

\begin{table} 
\caption{\label{tab:love} Linear and quadratic tidal constants computed for a polytropic equation of state $p = K \rho^{1+1/n}$. For each sampled value of $n$ we provide five sets of tidal constant between $b = 10^{-5}$ and $b = b_{\rm max}$, where $b := p(r=0)/\rho(r=0)$ is the ratio of central pressure to central density (which is used to parametrize the equilibrium sequence), and where $b_{\rm max}$ denotes the value at the configuration of maximum mass. The entries are equally spaced in $\log_{10} b$. A comparison between two independent sets of computations reveals agreement for all 6 significant digits.} 
\begin{ruledtabular}
\begin{tabular}{ccccccccc}
$n$ & $b$ & $M/R$ & $k_2$ & $k_4$ & $\ddot{k}_2$ & $\ddot{k}_4$ & $p_2$ & $p_4$ \\ 
\hline 
0.5 & 1.00000\e{-5} & 2.06441\e{-5} & 4.49096\e{-1}  & 1.25036\e{-1} & & & 1.17938 & 9.27857\e{-1} \\
& 1.70128\e{-4} & 3.50945\e{-4} & 4.48173\e{-1} & 1.24608\e{-1} & & & 1.17550 & 9.24205\e{-1} \\ 
& 2.89437\e{-3} & 5.89335\e{-3} & 4.32881\e{-1} & 1.17611\e{-1} & & & 1.11198 & 8.64710\e{-1} \\ 
& 4.92414\e{-2} & 8.19787\e{-2} & 2.57739\e{-1} & 5.00356\e{-2} & & & 4.80609\e{-1} & 3.21752\e{-1} \\ 
& 8.37735\e{-1} & 3.16668\e{-1} & 1.70183\e{-2} & 8.87795\e{-4} & & & 3.15134\e{-3} & 4.15926\e{-3} \\
& & & & & & \\
1.0 & 1.00000\e{-5} & 1.99989\e{-5} & 2.59872\e{-1} & 6.02278\e{-2} & 1.72580\e{-1} & 1.45070\e{-2} & 4.94857\e{-1} & 3.54684\e{-1} \\ 
& 1.34661\e{-4} & 2.69119\e{-4} & 2.59419\e{-1} & 6.00598\e{-2} & 1.72266\e{-1} & 1.44686\e{-2} & 4.93484\e{-1} & 3.53532\e{-1} \\ 
& 1.81337\e{-3} & 3.59016\e{-3} & 2.53428\e{-1} & 5.78571\e{-2} & 1.68112\e{-1} & 1.39644\e{-2} & 4.75471\e{-1} & 3.38467\e{-1} \\ 
& 2.44191\e{-2} & 4.28971\e{-2} & 1.89453\e{-1} & 3.64705\e{-2} & 1.23566\e{-1} & 8.90707\e{-3} & 2.98514\e{-1} & 1.97382\e{-1} \\ 
& 3.28830\e{-1} & 2.16430\e{-1} & 2.85515\e{-2} & 2.31968\e{-3} & 1.13698\e{-2} & 3.79990\e{-5} & 1.26251\e{-2} & 7.81802\e{-3} \\
& & & & & & \\
1.5 & 1.00000\e{-5} & 1.85691\e{-5} & 1.43257\e{-1} & 2.73867\e{-2} & 6.74485\e{-2} & 5.33396\e{-3} & 1.95218\e{-1} & 1.28910\e{-1} \\ 
& 1.10349\e{-4} & 2.04774\e{-4} & 1.43038\e{-1} & 2.73229\e{-2} & 6.73393\e{-2} & 5.32220\e{-3} & 1.94749\e{-1} & 1.28557\e{-1} \\ 
& 1.21768\e{-3} & 2.24345\e{-3} & 1.40650\e{-1} & 2.66327\e{-2} & 6.61507\e{-2} & 5.19472\e{-3} & 1.89671\e{-1} & 1.24751\e{-1} \\ 
& 1.34370\e{-2} & 2.29261\e{-2} & 1.17851\e{-1} & 2.03934\e{-2} & 5.48014\e{-2} & 4.02294\e{-3} & 1.43627\e{-1} & 9.11392\e{-2} \\ 
& 1.48275\e{-1} & 1.33622\e{-1} & 3.32914\e{-2} & 3.43752\e{-3} & 1.22637\e{-2} &  5.03042\e{-4} & 1.94309\e{-2} & 1.11044\e{-2} \\
& & & & & & \\
2.0 & 1.00000\e{-5} & 1.66157\e{-5} & 7.39258\e{-2} & 1.15050\e{-2} & 2.41628\e{-2} & 1.82433\e{-3} & 7.12689\e{-2} & 4.38190\e{-2} \\ 
& 9.21652\e{-5} & 1.53042\e{-4} & 7.38227\e{-2} & 1.14820\e{-2} & 2.41264\e{-2} & 1.82088\e{-3} & 7.11172\e{-2} & 4.37165\e{-2} \\ 
& 8.49442\e{-4} & 1.40241\e{-3} & 7.28823\e{-2} & 1.12735\e{-2} & 2.37952\e{-2} & 1.78948\e{-3} & 6.97390\e{-2} & 4.27863\e{-2} \\ 
& 7.82890\e{-3} & 1.22690\e{-2} & 6.49915\e{-2} & 9.57982\e{-3} & 2.10217\e{-2} & 1.53222\e{-3} & 5.85553\e{-2} & 3.53451\e{-2} \\ 
& 7.21552\e{-2} & 7.49332\e{-2} & 2.85166\e{-2} & 3.15909\e{-3} & 8.23899\e{-3} & 4.88879\e{-4} & 1.69290\e{-2} & 9.63285\e{-3} \\
& & & & & & \\
2.5 & 1.00000\e{-5} & 1.42934\e{-5} & 3.48455\e{-2} & 4.34038\e{-3} & 7.74388\e{-3} & 5.64580\e{-4} & 2.35777\e{-2} & 1.36223\e{-2} \\ 
& 7.23073\e{-5} & 1.03293\e{-4} & 3.48030\e{-2} & 4.33332\e{-3} & 7.73352\e{-3} & 5.63705\e{-4} & 2.35359\e{-2} & 1.35967\e{-2} \\ 
& 5.22834\e{-4} & 7.43820\e{-4} & 3.44979\e{-2} & 4.28278\e{-3} & 7.65915\e{-3} & 5.57433\e{-4} & 2.32363\e{-2} & 1.34134\e{-2} \\ 
& 3.78047\e{-3} & 5.22233\e{-3} & 3.24007\e{-2} & 3.94083\e{-3} & 7.14925\e{-3} & 5.14838\e{-4} & 2.12136\e{-2} & 1.21829\e{-2} \\ 
& 2.73356\e{-2} & 3.08993\e{-2} & 2.15146\e{-2} & 2.32536\e{-3} & 4.53405\e{-3} & 3.07954\e{-4} & 1.18096\e{-2} & 6.63746\e{-3}
\end{tabular} 
\end{ruledtabular} 
\end{table} 

Our numerical results for $p_2$ were previously displayed in Fig.~\ref{fig:p2}, and in Fig.~\ref{fig:p4} we present our results for $p_4$. To give context we also show in Fig.~\ref{fig:k4} the linear tidal constants $k_4$ and $\ddot{k}_4$ for the same polytropic models. The features are qualitatively the same as for the quadrupole sector, and we shall not repeat the detailed discussion presented in Sec.~\ref{intro:p2}. We provide a sample of numerical values in Table~\ref{tab:love}. The table also includes values for the dynamic tidal constants $\ddot{k}_2$ and $\ddot{k}_4$; these are copied from Ref.~\cite{pitre-poisson:24}, which did not calculate them for $n = 0.5$. We recall that $\ddot{k}_2$ plays an important role in the discussion of Sec.~\ref{intro:response}. 

\section{Tidal response}
\label{sec:response}

In this last section we fill in the technical steps that were omitted in the discussion of Sec.~\ref{intro:response}. We recall from Eq.~(\ref{Q_vs_E_intro}) that at leading order in the tidal interaction, a compact star's response to an applied  tidal field is measured by its mass quadrupole moment ${\cal Q}_{ab}(t)$, which is related to the tidal quadrupole moment ${\cal E}_{ab}(t)$ by 
\begin{equation}
G {\cal Q}_{ab} = -\frac{2}{3} k_2 R^5\, {\cal E}_{ab}
+ \frac{2}{3} \ddot{k}_2 \frac{R^8}{GM}\, \ddot{\cal E}_{ab}
+ \frac{2}{3} p_2 \frac{R^8}{GM}\, {\cal E}_{c\langle a} {\cal E}^c_{\ b\rangle}.
\label{Q_vs_E}
\end{equation}
This expression neglects higher-order terms in the simultaneous time-derivative and nonlinear expansion of the tidal field, and it also discards post-Newtonian corrections that were calculated in Ref.~\cite{poisson:21a}.

To turn Eq.~(\ref{Q_vs_E}) into something more concrete, we imagine that the star is in the presence of a companion of mass $M'$, and that the binary system is on a circular orbit of radius $r'$ with angular velocity $\Omega$. At the leading, Newtonian description of the orbit, this is given by
\begin{equation}
\Omega^2 = \frac{G(M+M')}{r^{\prime 3}}.
\label{Omega}
\end{equation}
We orient the coordinate system so that the orbit lies in the $x$-$y$ plane, and the unit vector
$n^a := ( \cos\Omega t, \sin\Omega t, 0 )$ points from the compact star to the companion. An orthogonal vector in the orbital plane is $l^a = (-\sin\Omega t, \cos\Omega t, 0)$. According to Eq.~(2.275) of Poisson and Will \cite{poisson-will:14}, the tidal moment is given by
\begin{equation}
{\cal E}_{ab} = -3 \frac{GM'}{r^{\prime 3}}\, n_{\langle ab \rangle},
\end{equation}
with $n_{\langle ab \rangle} := n_a n_b - \frac{1}{3} \delta_{ab}$. With $\dot{n}^a = \Omega\, l^a$ and $\ddot{n}^a = -\Omega^2\, n^a$, we find that
\begin{equation}
\ddot{\cal E}_{ab} = 6 \frac{GM'}{r^{\prime 3}} \Omega^2 \bigl( n_a n_b - l_a l_b \bigr)
\end{equation}
and
\begin{equation}
{\cal E}_{c\langle a} {\cal E}^c_{\ b\rangle} = 3 \frac{GM'}{r^{\prime 3}}
\frac{M'}{M+M'} \Omega^2\, n_{\langle ab \rangle}.
\end{equation}
In this last equation we choose to express the squared tidal moment in terms of $\Omega^2$ and the mass fraction $M'/(M+M')$. This serves to emphasize the fact that the time-derivative and quadratic terms in Eq.~(\ref{Q_vs_E}) are of the same order of magnitude; the dynamic and nonlinear terms in the relation between the mass and tidal moments are comparable to each other. 

We make the substitutions in Eq.~(\ref{Q_vs_E}) and calculate the tensorial expressions that involve the basis vectors $n^a$ and $l^a$. We find that the mass moment can be decomposed as
\begin{equation}
{\cal Q}_{ab} = {\cal Q}^{\rm static}_{ab} + {\cal Q}^{\rm cos}_{ab} + {\cal Q}^{\rm sin}_{ab},
\label{Q_decomposed}
\end{equation}
with
\begin{subequations}
\label{Q_pieces}
\begin{align}
{\cal Q}^{\rm static}_{ab} &= \frac{1}{3} \frac{M' R^5}{r^{\prime 3}}
\biggl( k_2 + p_2 \frac{M'}{M+M'} \frac{\Omega^2}{GM/R^3} \biggr)
\left(
\begin{array}{ccc}
  1 & 0 & 0 \\
  0 & 1 & 0 \\
  0 & 0 & -2
\end{array}
\right), \\
{\cal Q}^{\rm cos}_{ab} &= \frac{M' R^5}{r^{\prime 3}}
\biggl( k_2 + 4 \ddot{k}_2 \frac{\Omega^2}{GM/R^3}  
+ p_2 \frac{M'}{M+M'} \frac{\Omega^2}{GM/R^3} \biggr)
\left(
\begin{array}{ccc}
  1 & 0 & 0 \\
  0 & -1 & 0 \\
  0 & 0 & 0
\end{array}
\right) \cos 2\Omega t, \\
{\cal Q}^{\rm sin}_{ab} &= \frac{M' R^5}{r^{\prime 3}}
\biggl( k_2 + 4 \ddot{k}_2 \frac{\Omega^2}{GM/R^3}  
+ p_2 \frac{M'}{M+M'} \frac{\Omega^2}{GM/R^3} \biggr)
\left(
\begin{array}{ccc}
  0 & 1 & 0 \\
  1 & 0 & 0 \\
  0 & 0 & 0
\end{array}
\right) \sin 2\Omega t.
\end{align}
\end{subequations}
We see that the mass moment includes a static piece (which is constant in time) and time-dependent pieces that oscillate at the frequency $2\Omega$.

We now focus our attention on the oscillating pieces of the mass moment, and introduce the notation
\begin{equation}
\tilde{k}_2(2\Omega) := k_2 + 4 \ddot{k}_2 \frac{\Omega^2}{GM/R^3}  
+ p_2 \frac{M'}{M+M'} \frac{\Omega^2}{GM/R^3}
\label{ktilde1} 
\end{equation}
for the factor within brackets. The first term on the right is $k_2$, the linear and static tidal constant, and this contribution to $\tilde{k}_2(2\Omega)$ dominates during the regime of static tides. The term involving $\ddot{k}_2$ is a correction that accounts for the dynamic aspects of the tidal interaction, and the term proportional to $p_2$ is a nonlinear correction. Because both terms are proportional to $\Omega^2$, {\it it is natural to bring them together within an effective dynamical correction to the tidal response.} We do this in spite of the fact that the nonlinear term is actually of a static nature, as can be seen from the expression for ${\cal Q}^{\rm static}_{ab}$. Nevertheless, we maintain that to join these terms is both natural and compelling.

In the next stage of developments we follow our approach in Ref.~\cite{pitre-poisson:24} and pragmatically extend the domain of validity of Eq.~(\ref{ktilde1}) by writing it as
\begin{equation}
\tilde{k}_2(2\Omega) \simeq k_2 \biggl[ 1
- \frac{4 \ddot{k}_2 + p_2 \frac{M'}{M+M'}}{k_2} \frac{\Omega^2}{GM/R^3} \biggr]^{-1} 
\label{ktilde2} 
\end{equation}
and allowing the correction term proportional to $\Omega^2$ to become a substantial fraction of unity. This procedure was extensively motivated and justified in Ref.~\cite{pitre-poisson:24}. In the context of a Newtonian description of the tidal interaction, the repackaging of the response function was shown to be equivalent to a representation in terms of the star's normal modes of vibration, truncated to include but a single mode (the strongly dominant fundamental mode). Although inspired by the Newtonian mode description, the repackaging of Eq.~(\ref{ktilde2}) is formulated in general relativity and does not rely on a mode decomposition of the tidal deformation. As we argued in Ref.~\cite{pitre-poisson:24}, it is expected to provide a faithful description of the tidal response of a compact star.

The effective one-mode description of the tidal response is made more transparent when we rewrite Eq.~(\ref{ktilde2}) as we did back in Eq.~(\ref{ktilde3}). We refer the reader to Sec.~\ref{intro:response} for a complete discussion of this result.  

\begin{acknowledgments} 
We thank Sizheng Ma and Huan Yang for helpful discussions. This work was supported by the Natural Sciences and Engineering Research Council of Canada.  
\end{acknowledgments} 

\appendix

\section{Extrinsic curvature of the stellar surface}
\label{sec:extrinsic}

In this appendix we pick up where we left off at the end of Sec.~\ref{sec:junction} and compute $[K_{ab}]$, the jump in extrinsic curvature across the two faces of the boundary surface at $r = R + \delta R$. We recall that the extrinsic curvature is defined by Eq.~(\ref{h_K}). A laborious calculation returns
\begin{equation}
\bigl[ K_{tt} \bigr] = \biggl( 1 - \frac{2M}{R} \biggr)\, \Gamma, \qquad
\bigl[ K_{AB} \bigr] = R^2 \Omega_{AB}\, \Gamma
\label{ext_curvature} 
\end{equation}
for the nonvanishing components of the jump in extrinsic curvature across the boundary surface, with
\begin{equation}
\Gamma := 2\pi \frac{R^8}{M^3} \sqrt{1 - \frac{2M}{R}}\, \Bigl( \ovr{U}{1}{q}(r=R)\, \Eq \Bigr)^2
\, \lim \biggl( (\mu+p) \frac{d\mu}{dp} \biggr),  
\end{equation}
in which the limit of $(\mu+p) d\mu/dp$ is evaluated at the unperturbed surface, at which $r=R$. This limit vanishes for many equations of state, and for such cases we find that $[K_{ab}] = 0$ is satisfied. For other equations of state, however, the limit does not exist, and the junction conditions appear to be violated.

As an example, consider a polytropic equation of state with $p \propto \rho^{1+1/n}$, $\varepsilon = n p$, where $n$ is a constant. For such a fluid it can be shown that the energy density and pressure behave as $\mu \sim (R-r)^n$ and $p \sim (R-r)^{n-1}$, so that
\begin{equation}
(\mu + p) \frac{d\mu}{dp} \sim (R - r)^{n-1}.
\end{equation}
This vanishes at $r=R$ for $n > 1$, but it diverges when $n < 1$. (The special case $n=1$ must be handled separately, with the conclusion that the limit vanishes.) For polytropes with $n < 1$, therefore, the junction conditions $[K_{ab}] = 0$ do not seem to be satisfied. 

A nonvanishing $[K_{ab}]$ is usually associated with the presence of a thin layer of matter, with a surface energy-momentum tensor given by \cite{israel:66}
\begin{equation}
S_{ab} = -\frac{1}{8\pi} \Bigl( \bigl[ K_{ab} \bigr] - \bigl[ K \bigr] h_{ab} \Bigr),
\end{equation}
with $K := h^{ab} K_{ab}$ denoting the trace of the extrinsic curvature. With the components of Eq.~(\ref{ext_curvature}) we find that the surface tensor admits the decomposition
\begin{equation}
S^{ab} = \sigma\, u^a u^b
\end{equation}
of a pressureless fluid, in terms of a surface density $\sigma = -\Gamma/(4\pi)$ and a normalized velocity field with nonvanishing component $u^t = (1-2M/R)^{-1/2}$. In this case, however, the interpretation is dubious because $\Gamma$ is formally infinite when it does not vanish, and the surface density turns out to be infinitely negative. 

The apparent inability to enforce $[K_{ab}] = 0$ for some equations of state is a curious outcome. We believe that this is an artifice of our second-order perturbative analysis: While $[K_{ab}]$ is actually regular (and vanishing) on the deformed boundary surface, it becomes problematic when we express it as an expansion in powers of $\delta R$. We argue this point by exhibiting a simpler manifestation of the same phenomenon. Consider any equation of state such that $\mu = 0$ when $p=0$ and $h = 1$. For such an equation of state, it is unambiguous that the perturbed density vanishes on the perturbed surface, $(\mu+\delta \mu)(r = R + \delta R) = 0$. Expressing this as a perturbative expansion, however, produces an ambiguity. The expansion, carried out through first order in $\delta R$, is given by
\begin{align}
(\mu+\delta \mu)(r = R + \delta R) &=
\mu(r=R) + \frac{d\mu}{dr} \biggr|_{r=R} \delta R + \delta \mu(r=R)
\nonumber \\ 
&= \mu(r=R) + \frac{d\mu}{dp} \biggr|_{r=R} \Bigl( p'(r=R)\, \delta R + \delta p (r=R) \Bigr).
\end{align}
The first term vanishes, and in the second term, the factor within brackets vanishes also; but it is multiplied by $d\mu/dp$, which diverges when $r \to R$ for some equations of state, making the right-hand side ill defined. The perturbative expansion can therefore drive a well-behaved quantity to delinquency by using the unperturbed surface as a reference position; fluid variables and gravitational potentials have a limited degree of differentiability there. 

The variables $\ovr{a}{j}{\ell}$ and $\ovr{b}{j}{\ell}$ introduced in Sec.~\ref{sec:pert_eqns} permit the numerical integration of the perturbation equations for any equation of state, and the junction conditions of Eq.~(\ref{junction}) form a necessary and sufficient set to identify a unique solution to the global problem. We shall not be concerned with the formal difficulties that arise when the $r \to R$ limit of $(\mu+p) d\mu/dp$ does not exist.

\bibliography{nonlinear}

\begin{thebibliography}{40}%
\makeatletter
\providecommand \@ifxundefined [1]{%
 \@ifx{#1\undefined}
}%
\providecommand \@ifnum [1]{%
 \ifnum #1\expandafter \@firstoftwo
 \else \expandafter \@secondoftwo
 \fi
}%
\providecommand \@ifx [1]{%
 \ifx #1\expandafter \@firstoftwo
 \else \expandafter \@secondoftwo
 \fi
}%
\providecommand \natexlab [1]{#1}%
\providecommand \enquote  [1]{``#1''}%
\providecommand \bibnamefont  [1]{#1}%
\providecommand \bibfnamefont [1]{#1}%
\providecommand \citenamefont [1]{#1}%
\providecommand \href@noop [0]{\@secondoftwo}%
\providecommand \href [0]{\begingroup \@sanitize@url \@href}%
\providecommand \@href[1]{\@@startlink{#1}\@@href}%
\providecommand \@@href[1]{\endgroup#1\@@endlink}%
\providecommand \@sanitize@url [0]{\catcode `\\12\catcode `\$12\catcode
  `\&12\catcode `\#12\catcode `\^12\catcode `\_12\catcode `\%12\relax}%
\providecommand \@@startlink[1]{}%
\providecommand \@@endlink[0]{}%
\providecommand \url  [0]{\begingroup\@sanitize@url \@url }%
\providecommand \@url [1]{\endgroup\@href {#1}{\urlprefix }}%
\providecommand \urlprefix  [0]{URL }%
\providecommand \Eprint [0]{\href }%
\providecommand \doibase [0]{https://doi.org/}%
\providecommand \selectlanguage [0]{\@gobble}%
\providecommand \bibinfo  [0]{\@secondoftwo}%
\providecommand \bibfield  [0]{\@secondoftwo}%
\providecommand \translation [1]{[#1]}%
\providecommand \BibitemOpen [0]{}%
\providecommand \bibitemStop [0]{}%
\providecommand \bibitemNoStop [0]{.\EOS\space}%
\providecommand \EOS [0]{\spacefactor3000\relax}%
\providecommand \BibitemShut  [1]{\csname bibitem#1\endcsname}%
\let\auto@bib@innerbib\@empty
\bibitem [{\citenamefont {Flanagan}\ and\ \citenamefont
  {Hinderer}(2008)}]{flanagan-hinderer:08}%
  \BibitemOpen
  \bibfield  {author} {\bibinfo {author} {\bibfnamefont {E.~E.}\ \bibnamefont
  {Flanagan}}\ and\ \bibinfo {author} {\bibfnamefont {T.}~\bibnamefont
  {Hinderer}},\ }\bibfield  {title} {\bibinfo {title} {{Constraining
  neutron-star tidal Love numbers with gravitational-wave detectors}},\ }\href
  {https://doi.org/10.1103/PhysRevD.77.021502} {\bibfield  {journal} {\bibinfo
  {journal} {Phys. Rev. D}\ }\textbf {\bibinfo {volume} {77}},\ \bibinfo
  {pages} {021502} (\bibinfo {year} {2008})}\BibitemShut {NoStop}%
\bibitem [{\citenamefont {\"{O}zel}\ and\ \citenamefont
  {Freire}(2016)}]{ozel-freire:16}%
  \BibitemOpen
  \bibfield  {author} {\bibinfo {author} {\bibfnamefont {F.}~\bibnamefont
  {\"{O}zel}}\ and\ \bibinfo {author} {\bibfnamefont {P.}~\bibnamefont
  {Freire}},\ }\bibfield  {title} {\bibinfo {title} {{Masses, radii, and the
  equation of state of neutron stars}},\ }\href
  {https://doi.org/10.1146/annurev-astro-081915-023322} {\bibfield  {journal}
  {\bibinfo  {journal} {Ann. Rev. Astron. Astrophys.}\ }\textbf {\bibinfo
  {volume} {54}},\ \bibinfo {pages} {401} (\bibinfo {year} {2016})}\BibitemShut
  {NoStop}%
\bibitem [{\citenamefont {Oertel}\ \emph {et~al.}(2017)\citenamefont {Oertel},
  \citenamefont {Hempel}, \citenamefont {Kl\"ahn},\ and\ \citenamefont
  {Typel}}]{oertel-etal:17}%
  \BibitemOpen
  \bibfield  {author} {\bibinfo {author} {\bibfnamefont {M.}~\bibnamefont
  {Oertel}}, \bibinfo {author} {\bibfnamefont {M.}~\bibnamefont {Hempel}},
  \bibinfo {author} {\bibfnamefont {T.}~\bibnamefont {Kl\"ahn}},\ and\ \bibinfo
  {author} {\bibfnamefont {S.}~\bibnamefont {Typel}},\ }\bibfield  {title}
  {\bibinfo {title} {{Equations of state for supernovae and compact stars}},\
  }\href {https://doi.org/10.1103/RevModPhys.89.015007} {\bibfield  {journal}
  {\bibinfo  {journal} {Rev. Mod. Phys.}\ }\textbf {\bibinfo {volume} {89}},\
  \bibinfo {pages} {015007} (\bibinfo {year} {2017})}\BibitemShut {NoStop}%
\bibitem [{\citenamefont {Baym}\ \emph {et~al.}(2018)\citenamefont {Baym},
  \citenamefont {Hatsuda}, \citenamefont {Kojo}, \citenamefont {Powell},
  \citenamefont {Song},\ and\ \citenamefont {Takatsuka}}]{baym-etal:18}%
  \BibitemOpen
  \bibfield  {author} {\bibinfo {author} {\bibfnamefont {G.}~\bibnamefont
  {Baym}}, \bibinfo {author} {\bibfnamefont {T.}~\bibnamefont {Hatsuda}},
  \bibinfo {author} {\bibfnamefont {T.}~\bibnamefont {Kojo}}, \bibinfo {author}
  {\bibfnamefont {P.~D.}\ \bibnamefont {Powell}}, \bibinfo {author}
  {\bibfnamefont {Y.}~\bibnamefont {Song}},\ and\ \bibinfo {author}
  {\bibfnamefont {T.}~\bibnamefont {Takatsuka}},\ }\bibfield  {title} {\bibinfo
  {title} {{From hadrons to quarks in neutron stars: a review}},\ }\href@noop
  {} {\bibfield  {journal} {\bibinfo  {journal} {Rept. Prog. Phys.}\ }\textbf
  {\bibinfo {volume} {81}},\ \bibinfo {pages} {056902} (\bibinfo {year}
  {2018})}\BibitemShut {NoStop}%
\bibitem [{\citenamefont {Abbott}\ \emph {et~al.}(2017)\citenamefont {Abbott}
  \emph {et~al.}}]{GW170817:17}%
  \BibitemOpen
  \bibfield  {author} {\bibinfo {author} {\bibfnamefont {B.~P.}\ \bibnamefont
  {Abbott}} \emph {et~al.} (\bibinfo {collaboration} {LIGO Scientific
  Collaboration and Virgo Collaboration}),\ }\bibfield  {title} {\bibinfo
  {title} {{GW170817: Observation of Gravitational Waves from a Binary Neutron
  Star Inspiral}},\ }\href {https://doi.org/10.1103/PhysRevLett.119.161101}
  {\bibfield  {journal} {\bibinfo  {journal} {Phys. Rev. Lett.}\ }\textbf
  {\bibinfo {volume} {119}},\ \bibinfo {pages} {161101} (\bibinfo {year}
  {2017})}\BibitemShut {NoStop}%
\bibitem [{\citenamefont {Abbott}\ \emph {et~al.}(2018)\citenamefont {Abbott}
  \emph {et~al.}}]{GW170817:18}%
  \BibitemOpen
  \bibfield  {author} {\bibinfo {author} {\bibfnamefont {B.~P.}\ \bibnamefont
  {Abbott}} \emph {et~al.} (\bibinfo {collaboration} {The LIGO Scientific
  Collaboration and the Virgo Collaboration}),\ }\bibfield  {title} {\bibinfo
  {title} {{GW170817: Measurements of Neutron Star Radii and Equation of
  State}},\ }\href {https://doi.org/10.1103/PhysRevLett.121.161101} {\bibfield
  {journal} {\bibinfo  {journal} {Phys. Rev. Lett.}\ }\textbf {\bibinfo
  {volume} {121}},\ \bibinfo {pages} {161101} (\bibinfo {year}
  {2018})}\BibitemShut {NoStop}%
\bibitem [{\citenamefont {Narikawa}\ \emph {et~al.}(2021)\citenamefont
  {Narikawa}, \citenamefont {Uchikata},\ and\ \citenamefont
  {Tanaka}}]{narikawa-uchikata-tanaka:21}%
  \BibitemOpen
  \bibfield  {author} {\bibinfo {author} {\bibfnamefont {T.}~\bibnamefont
  {Narikawa}}, \bibinfo {author} {\bibfnamefont {N.}~\bibnamefont {Uchikata}},\
  and\ \bibinfo {author} {\bibfnamefont {T.}~\bibnamefont {Tanaka}},\
  }\bibfield  {title} {\bibinfo {title} {{Gravitational-wave constraints on the
  GWTC-2 events by measuring the tidal deformability and the spin-induced
  quadrupole moment}},\ }\href {https://doi.org/10.1103/PhysRevD.104.084056}
  {\bibfield  {journal} {\bibinfo  {journal} {Phys. Rev. D}\ }\textbf {\bibinfo
  {volume} {104}},\ \bibinfo {pages} {084056} (\bibinfo {year}
  {2021})}\BibitemShut {NoStop}%
\bibitem [{\citenamefont {Landry}\ \emph {et~al.}(2020)\citenamefont {Landry},
  \citenamefont {Essick},\ and\ \citenamefont
  {Chatziioannou}}]{landry-essick-reed-chatziioannou:20}%
  \BibitemOpen
  \bibfield  {author} {\bibinfo {author} {\bibfnamefont {P.}~\bibnamefont
  {Landry}}, \bibinfo {author} {\bibfnamefont {R.}~\bibnamefont {Essick}},\
  and\ \bibinfo {author} {\bibfnamefont {K.}~\bibnamefont {Chatziioannou}},\
  }\bibfield  {title} {\bibinfo {title} {{Nonparametric constraints on neutron
  star matter with existing and upcoming gravitational wave and pulsar
  observations}},\ }\href {https://doi.org/10.1103/PhysRevD.101.123007}
  {\bibfield  {journal} {\bibinfo  {journal} {Phys. Rev. D}\ }\textbf {\bibinfo
  {volume} {101}},\ \bibinfo {pages} {123007} (\bibinfo {year}
  {2020})}\BibitemShut {NoStop}%
\bibitem [{\citenamefont {Chatziioannou}(2020)}]{chatziioannou:20}%
  \BibitemOpen
  \bibfield  {author} {\bibinfo {author} {\bibfnamefont {K.}~\bibnamefont
  {Chatziioannou}},\ }\bibfield  {title} {\bibinfo {title} {{Neutron star tidal
  deformability and equation-of-state constraints}},\ }\href
  {https://doi.org/https://doi.org/10.1007/s10714-020-02754-3} {\bibfield
  {journal} {\bibinfo  {journal} {Gen. Rel. Grav.}\ }\textbf {\bibinfo {volume}
  {52}},\ \bibinfo {pages} {109} (\bibinfo {year} {2020})}\BibitemShut
  {NoStop}%
\bibitem [{\citenamefont {Pacilio}\ \emph {et~al.}(2022)\citenamefont
  {Pacilio}, \citenamefont {Maselli}, \citenamefont {Fasano},\ and\
  \citenamefont {Pani}}]{pacilio-maselli-fasano-pani:22}%
  \BibitemOpen
  \bibfield  {author} {\bibinfo {author} {\bibfnamefont {C.}~\bibnamefont
  {Pacilio}}, \bibinfo {author} {\bibfnamefont {A.}~\bibnamefont {Maselli}},
  \bibinfo {author} {\bibfnamefont {M.}~\bibnamefont {Fasano}},\ and\ \bibinfo
  {author} {\bibfnamefont {P.}~\bibnamefont {Pani}},\ }\bibfield  {title}
  {\bibinfo {title} {{Ranking Love Numbers for the Neutron Star Equation of
  State: The Need for Third-Generation Detectors}},\ }\href
  {https://doi.org/10.1103/PhysRevLett.128.101101} {\bibfield  {journal}
  {\bibinfo  {journal} {Phys. Rev. Lett.}\ }\textbf {\bibinfo {volume} {128}},\
  \bibinfo {pages} {101101} (\bibinfo {year} {2022})}\BibitemShut {NoStop}%
\bibitem [{\citenamefont {Hinderer}\ \emph {et~al.}(2016)\citenamefont
  {Hinderer}, \citenamefont {Taracchini}, \citenamefont {Foucart},
  \citenamefont {Buonanno}, \citenamefont {Steinhoff}, \citenamefont {Duez},
  \citenamefont {Kidder}, \citenamefont {Pfeiffer}, \citenamefont {Scheel},
  \citenamefont {Szilagyi}, \citenamefont {Hotokezaka}, \citenamefont
  {Kyutoku}, \citenamefont {Shibata},\ and\ \citenamefont
  {Carpenter}}]{hinderer-etal:16}%
  \BibitemOpen
  \bibfield  {author} {\bibinfo {author} {\bibfnamefont {T.}~\bibnamefont
  {Hinderer}}, \bibinfo {author} {\bibfnamefont {A.}~\bibnamefont
  {Taracchini}}, \bibinfo {author} {\bibfnamefont {F.}~\bibnamefont {Foucart}},
  \bibinfo {author} {\bibfnamefont {A.}~\bibnamefont {Buonanno}}, \bibinfo
  {author} {\bibfnamefont {J.}~\bibnamefont {Steinhoff}}, \bibinfo {author}
  {\bibfnamefont {M.}~\bibnamefont {Duez}}, \bibinfo {author} {\bibfnamefont
  {L.~E.}\ \bibnamefont {Kidder}}, \bibinfo {author} {\bibfnamefont {H.~P.}\
  \bibnamefont {Pfeiffer}}, \bibinfo {author} {\bibfnamefont {M.~A.}\
  \bibnamefont {Scheel}}, \bibinfo {author} {\bibfnamefont {B.}~\bibnamefont
  {Szilagyi}}, \bibinfo {author} {\bibfnamefont {K.}~\bibnamefont
  {Hotokezaka}}, \bibinfo {author} {\bibfnamefont {K.}~\bibnamefont {Kyutoku}},
  \bibinfo {author} {\bibfnamefont {M.}~\bibnamefont {Shibata}},\ and\ \bibinfo
  {author} {\bibfnamefont {C.~W.}\ \bibnamefont {Carpenter}},\ }\bibfield
  {title} {\bibinfo {title} {{Effects of Neutron-Star Dynamic Tides on
  Gravitational Waveforms within the Effective-One-Body Approach}},\ }\href
  {https://doi.org/10.1103/PhysRevLett.116.181101} {\bibfield  {journal}
  {\bibinfo  {journal} {Phys. Rev. Lett.}\ }\textbf {\bibinfo {volume} {116}},\
  \bibinfo {pages} {181101} (\bibinfo {year} {2016})}\BibitemShut {NoStop}%
\bibitem [{\citenamefont {Steinhoff}\ \emph {et~al.}(2016)\citenamefont
  {Steinhoff}, \citenamefont {Hinderer}, \citenamefont {Buonanno},\ and\
  \citenamefont {Taracchini}}]{steinhoff-etal:16}%
  \BibitemOpen
  \bibfield  {author} {\bibinfo {author} {\bibfnamefont {J.}~\bibnamefont
  {Steinhoff}}, \bibinfo {author} {\bibfnamefont {T.}~\bibnamefont {Hinderer}},
  \bibinfo {author} {\bibfnamefont {A.}~\bibnamefont {Buonanno}},\ and\
  \bibinfo {author} {\bibfnamefont {A.}~\bibnamefont {Taracchini}},\ }\bibfield
   {title} {\bibinfo {title} {{Dynamical tides in general relativity: Effective
  action and effective-one-body Hamiltonian}},\ }\href
  {https://doi.org/10.1103/PhysRevD.94.104028} {\bibfield  {journal} {\bibinfo
  {journal} {Phys. Rev. D}\ }\textbf {\bibinfo {volume} {94}},\ \bibinfo
  {pages} {104028} (\bibinfo {year} {2016})}\BibitemShut {NoStop}%
\bibitem [{\citenamefont {Steinhoff}\ \emph {et~al.}(2021)\citenamefont
  {Steinhoff}, \citenamefont {Hinderer}, \citenamefont {Dietrich},\ and\
  \citenamefont {Foucart}}]{steinhoff-etal:21}%
  \BibitemOpen
  \bibfield  {author} {\bibinfo {author} {\bibfnamefont {J.}~\bibnamefont
  {Steinhoff}}, \bibinfo {author} {\bibfnamefont {T.}~\bibnamefont {Hinderer}},
  \bibinfo {author} {\bibfnamefont {T.}~\bibnamefont {Dietrich}},\ and\
  \bibinfo {author} {\bibfnamefont {F.}~\bibnamefont {Foucart}},\ }\bibfield
  {title} {\bibinfo {title} {{Spin effects on neutron star fundamental-mode
  dynamical tides: Phenomenology and comparison to numerical simulations}},\
  }\href {https://doi.org/10.1103/PhysRevResearch.3.033129} {\bibfield
  {journal} {\bibinfo  {journal} {Phys. Rev. Res.}\ }\textbf {\bibinfo {volume}
  {3}},\ \bibinfo {pages} {033129} (\bibinfo {year} {2021})}\BibitemShut
  {NoStop}%
\bibitem [{\citenamefont {Andersson}\ and\ \citenamefont
  {Pnigouras}(2021)}]{andersson-pnigouras:21}%
  \BibitemOpen
  \bibfield  {author} {\bibinfo {author} {\bibfnamefont {N.}~\bibnamefont
  {Andersson}}\ and\ \bibinfo {author} {\bibfnamefont {P.}~\bibnamefont
  {Pnigouras}},\ }\bibfield  {title} {\bibinfo {title} {{The phenomenology of
  dynamical neutron star tides}},\ }\href
  {https://doi.org/10.1093/mnras/stab371} {\bibfield  {journal} {\bibinfo
  {journal} {M.N.R.A.S.}\ }\textbf {\bibinfo {volume} {503}},\ \bibinfo {pages}
  {533} (\bibinfo {year} {2021})}\BibitemShut {NoStop}%
\bibitem [{\citenamefont {Passamonti}\ \emph {et~al.}(2021)\citenamefont
  {Passamonti}, \citenamefont {Andersson},\ and\ \citenamefont
  {Pnigouras}}]{passamonti-andersson-pnigouras:21}%
  \BibitemOpen
  \bibfield  {author} {\bibinfo {author} {\bibfnamefont {A.}~\bibnamefont
  {Passamonti}}, \bibinfo {author} {\bibfnamefont {N.}~\bibnamefont
  {Andersson}},\ and\ \bibinfo {author} {\bibfnamefont {P.}~\bibnamefont
  {Pnigouras}},\ }\bibfield  {title} {\bibinfo {title} {{Dynamical tides in
  neutron stars: the impact of the crust}},\ }\href
  {https://doi.org/10.1093/mnras/stab870} {\bibfield  {journal} {\bibinfo
  {journal} {M.N.R.A.S.}\ }\textbf {\bibinfo {volume} {504}},\ \bibinfo {pages}
  {1273} (\bibinfo {year} {2021})}\BibitemShut {NoStop}%
\bibitem [{\citenamefont {Passamonti}\ \emph {et~al.}(2022)\citenamefont
  {Passamonti}, \citenamefont {Andersson},\ and\ \citenamefont
  {Pnigouras}}]{passamonti-andersson-pnigouras:22}%
  \BibitemOpen
  \bibfield  {author} {\bibinfo {author} {\bibfnamefont {A.}~\bibnamefont
  {Passamonti}}, \bibinfo {author} {\bibfnamefont {N.}~\bibnamefont
  {Andersson}},\ and\ \bibinfo {author} {\bibfnamefont {P.}~\bibnamefont
  {Pnigouras}},\ }\bibfield  {title} {\bibinfo {title} {{Dynamical tides in
  superfluid neutron stars}},\ }\href {https://doi.org/10.1093/mnras/stac1380}
  {\bibfield  {journal} {\bibinfo  {journal} {M.N.R.A.S.}\ }\textbf {\bibinfo
  {volume} {514}},\ \bibinfo {pages} {1494} (\bibinfo {year}
  {2022})}\BibitemShut {NoStop}%
\bibitem [{\citenamefont {Schmidt}\ and\ \citenamefont
  {Hinderer}(2019)}]{schmidt-hinderer:19}%
  \BibitemOpen
  \bibfield  {author} {\bibinfo {author} {\bibfnamefont {P.}~\bibnamefont
  {Schmidt}}\ and\ \bibinfo {author} {\bibfnamefont {T.}~\bibnamefont
  {Hinderer}},\ }\bibfield  {title} {\bibinfo {title} {{Frequency domain model
  of $f$-mode dynamic tides in gravitational waveforms from compact binary
  inspirals}},\ }\href {https://doi.org/10.1103/PhysRevD.100.021501} {\bibfield
   {journal} {\bibinfo  {journal} {Phys. Rev. D}\ }\textbf {\bibinfo {volume}
  {100}},\ \bibinfo {pages} {021501} (\bibinfo {year} {2019})}\BibitemShut
  {NoStop}%
\bibitem [{\citenamefont {Mandal}\ \emph {et~al.}(2023)\citenamefont {Mandal},
  \citenamefont {Mastrolia}, \citenamefont {Silva}, \citenamefont {Patil},\
  and\ \citenamefont {Steinhoff}}]{mandal-etal:23}%
  \BibitemOpen
  \bibfield  {author} {\bibinfo {author} {\bibfnamefont {M.~K.}\ \bibnamefont
  {Mandal}}, \bibinfo {author} {\bibfnamefont {P.}~\bibnamefont {Mastrolia}},
  \bibinfo {author} {\bibfnamefont {H.~O.}\ \bibnamefont {Silva}}, \bibinfo
  {author} {\bibfnamefont {R.}~\bibnamefont {Patil}},\ and\ \bibinfo {author}
  {\bibfnamefont {J.}~\bibnamefont {Steinhoff}},\ }\bibfield  {title} {\bibinfo
  {title} {{Gravitoelectric dynamical tides at second post-Newtonian order}},\
  }\href {https://doi.org/10.1007/JHEP11(2023)067} {\bibfield  {journal}
  {\bibinfo  {journal} {J. High. Energ. Phys.}\ }\textbf {\bibinfo {volume}
  {2023}},\ \bibinfo {pages} {67} (\bibinfo {year} {2023})}\BibitemShut
  {NoStop}%
\bibitem [{\citenamefont {Mandal}\ \emph {et~al.}(2024)\citenamefont {Mandal},
  \citenamefont {Mastrolia}, \citenamefont {Silva}, \citenamefont {Patil},\
  and\ \citenamefont {Steinhoff}}]{mandal-etal:24}%
  \BibitemOpen
  \bibfield  {author} {\bibinfo {author} {\bibfnamefont {M.~K.}\ \bibnamefont
  {Mandal}}, \bibinfo {author} {\bibfnamefont {P.}~\bibnamefont {Mastrolia}},
  \bibinfo {author} {\bibfnamefont {H.~O.}\ \bibnamefont {Silva}}, \bibinfo
  {author} {\bibfnamefont {R.}~\bibnamefont {Patil}},\ and\ \bibinfo {author}
  {\bibfnamefont {J.}~\bibnamefont {Steinhoff}},\ }\bibfield  {title} {\bibinfo
  {title} {{Renormalizing Love: tidal effects at the third post-Newtonian
  order}},\ }\href {https://doi.org/10.1007/JHEP02(2024)188} {\bibfield
  {journal} {\bibinfo  {journal} {J. High. Energ. Phys.}\ }\textbf {\bibinfo
  {volume} {2024}},\ \bibinfo {pages} {188} (\bibinfo {year}
  {2024})}\BibitemShut {NoStop}%
\bibitem [{\citenamefont {Andersson}\ and\ \citenamefont
  {Ho}(2018)}]{andersson-ho:18}%
  \BibitemOpen
  \bibfield  {author} {\bibinfo {author} {\bibfnamefont {N.}~\bibnamefont
  {Andersson}}\ and\ \bibinfo {author} {\bibfnamefont {W.~C.~G.}\ \bibnamefont
  {Ho}},\ }\bibfield  {title} {\bibinfo {title} {{Using gravitational-wave data
  to constrain dynamical tides in neutron star binaries}},\ }\href
  {https://doi.org/10.1103/PhysRevD.97.023016} {\bibfield  {journal} {\bibinfo
  {journal} {Phys. Rev. D}\ }\textbf {\bibinfo {volume} {97}},\ \bibinfo
  {pages} {023016} (\bibinfo {year} {2018})}\BibitemShut {NoStop}%
\bibitem [{\citenamefont {Williams}\ \emph {et~al.}(2022)\citenamefont
  {Williams}, \citenamefont {Pratten},\ and\ \citenamefont
  {Schmidt}}]{williams-pratten-schmidt:22}%
  \BibitemOpen
  \bibfield  {author} {\bibinfo {author} {\bibfnamefont {N.}~\bibnamefont
  {Williams}}, \bibinfo {author} {\bibfnamefont {G.}~\bibnamefont {Pratten}},\
  and\ \bibinfo {author} {\bibfnamefont {P.}~\bibnamefont {Schmidt}},\
  }\bibfield  {title} {\bibinfo {title} {{Prospects for distinguishing
  dynamical tides in inspiralling binary neutron stars with third generation
  gravitational-wave detectors}},\ }\href
  {https://doi.org/10.1103/PhysRevD.105.123032} {\bibfield  {journal} {\bibinfo
   {journal} {Phys. Rev. D}\ }\textbf {\bibinfo {volume} {105}},\ \bibinfo
  {pages} {123032} (\bibinfo {year} {2022})}\BibitemShut {NoStop}%
\bibitem [{\citenamefont {Pratten}\ \emph {et~al.}(2022)\citenamefont
  {Pratten}, \citenamefont {Schmidt},\ and\ \citenamefont
  {Williams}}]{pratten-schmidt-williams:22}%
  \BibitemOpen
  \bibfield  {author} {\bibinfo {author} {\bibfnamefont {G.}~\bibnamefont
  {Pratten}}, \bibinfo {author} {\bibfnamefont {P.}~\bibnamefont {Schmidt}},\
  and\ \bibinfo {author} {\bibfnamefont {N.}~\bibnamefont {Williams}},\
  }\bibfield  {title} {\bibinfo {title} {{Impact of Dynamical Tides on the
  Reconstruction of the Neutron Star Equation of State}},\ }\href
  {https://doi.org/10.1103/PhysRevLett.129.081102} {\bibfield  {journal}
  {\bibinfo  {journal} {Phys. Rev. Lett.}\ }\textbf {\bibinfo {volume} {129}},\
  \bibinfo {pages} {081102} (\bibinfo {year} {2022})}\BibitemShut {NoStop}%
\bibitem [{\citenamefont {Yu}\ \emph {et~al.}(2022)\citenamefont {Yu},
  \citenamefont {Weinberg}, \citenamefont {Arras}, \citenamefont {Kwon},\ and\
  \citenamefont {Venumadhav}}]{yu-etal:22}%
  \BibitemOpen
  \bibfield  {author} {\bibinfo {author} {\bibfnamefont {H.}~\bibnamefont
  {Yu}}, \bibinfo {author} {\bibfnamefont {N.~N.}\ \bibnamefont {Weinberg}},
  \bibinfo {author} {\bibfnamefont {P.}~\bibnamefont {Arras}}, \bibinfo
  {author} {\bibfnamefont {J.}~\bibnamefont {Kwon}},\ and\ \bibinfo {author}
  {\bibfnamefont {T.}~\bibnamefont {Venumadhav}},\ }\bibfield  {title}
  {\bibinfo {title} {{Beyond the linear tide: impact of the non-linear tidal
  response of neutron stars on gravitational waveforms from binary
  inspirals}},\ }\href {https://doi.org/10.1093/mnras/stac3614} {\bibfield
  {journal} {\bibinfo  {journal} {M.N.R.A.S.}\ }\textbf {\bibinfo {volume}
  {519}},\ \bibinfo {pages} {4325} (\bibinfo {year} {2022})}\BibitemShut
  {NoStop}%
\bibitem [{\citenamefont {Weinberg}\ \emph {et~al.}(2012)\citenamefont
  {Weinberg}, \citenamefont {Arras}, \citenamefont {Quataert},\ and\
  \citenamefont {Burkart}}]{weinberg-etal:12}%
  \BibitemOpen
  \bibfield  {author} {\bibinfo {author} {\bibfnamefont {N.~N.}\ \bibnamefont
  {Weinberg}}, \bibinfo {author} {\bibfnamefont {P.}~\bibnamefont {Arras}},
  \bibinfo {author} {\bibfnamefont {E.}~\bibnamefont {Quataert}},\ and\
  \bibinfo {author} {\bibfnamefont {J.}~\bibnamefont {Burkart}},\ }\bibfield
  {title} {\bibinfo {title} {{Nonlinear tides in close binary systems}},\
  }\href {https://doi.org/doi:10.1088/0004-637X/751/2/136} {\bibfield
  {journal} {\bibinfo  {journal} {Astrophys. J.}\ }\textbf {\bibinfo {volume}
  {751}},\ \bibinfo {pages} {136} (\bibinfo {year} {2012})}\BibitemShut
  {NoStop}%
\bibitem [{\citenamefont {Schenk}\ \emph {et~al.}(2001)\citenamefont {Schenk},
  \citenamefont {Arras}, \citenamefont {Flanagan}, \citenamefont {Teukolsky},\
  and\ \citenamefont {Wasserman}}]{schenk-etal:01}%
  \BibitemOpen
  \bibfield  {author} {\bibinfo {author} {\bibfnamefont {A.~K.}\ \bibnamefont
  {Schenk}}, \bibinfo {author} {\bibfnamefont {P.}~\bibnamefont {Arras}},
  \bibinfo {author} {\bibfnamefont {E.~E.}\ \bibnamefont {Flanagan}}, \bibinfo
  {author} {\bibfnamefont {S.~A.}\ \bibnamefont {Teukolsky}},\ and\ \bibinfo
  {author} {\bibfnamefont {I.}~\bibnamefont {Wasserman}},\ }\bibfield  {title}
  {\bibinfo {title} {{Nonlinear mode coupling in rotating stars and the r-mode
  instability in neutron stars}},\ }\href
  {https://doi.org/10.1103/PhysRevD.65.024001} {\bibfield  {journal} {\bibinfo
  {journal} {Phys. Rev. D}\ }\textbf {\bibinfo {volume} {65}},\ \bibinfo
  {pages} {024001} (\bibinfo {year} {2001})}\BibitemShut {NoStop}%
\bibitem [{\citenamefont {Hossein~Nouri}\ \emph {et~al.}(2022)\citenamefont
  {Hossein~Nouri}, \citenamefont {Bose}, \citenamefont {Duez},\ and\
  \citenamefont {Das}}]{nouri-etal:22}%
  \BibitemOpen
  \bibfield  {author} {\bibinfo {author} {\bibfnamefont {F.}~\bibnamefont
  {Hossein~Nouri}}, \bibinfo {author} {\bibfnamefont {S.}~\bibnamefont {Bose}},
  \bibinfo {author} {\bibfnamefont {M.~D.}\ \bibnamefont {Duez}},\ and\
  \bibinfo {author} {\bibfnamefont {A.}~\bibnamefont {Das}},\ }\bibfield
  {title} {\bibinfo {title} {{Nonlinear mode-tide coupling in coalescing binary
  neutron stars with relativistic corrections}},\ }\href
  {https://doi.org/10.1103/PhysRevD.106.083001} {\bibfield  {journal} {\bibinfo
   {journal} {Phys. Rev. D}\ }\textbf {\bibinfo {volume} {106}},\ \bibinfo
  {pages} {083001} (\bibinfo {year} {2022})}\BibitemShut {NoStop}%
\bibitem [{\citenamefont {Kuan}\ \emph {et~al.}(2024)\citenamefont {Kuan},
  \citenamefont {Kiuchi},\ and\ \citenamefont
  {Shibata}}]{kuan-kiuchi-shibata:24}%
  \BibitemOpen
  \bibfield  {author} {\bibinfo {author} {\bibfnamefont {H.-J.}\ \bibnamefont
  {Kuan}}, \bibinfo {author} {\bibfnamefont {K.}~\bibnamefont {Kiuchi}},\ and\
  \bibinfo {author} {\bibfnamefont {M.}~\bibnamefont {Shibata}},\ }\href@noop
  {} {\bibinfo {title} {{Tidal resonance in binary neutron star inspirals: A
  high-precision study in numerical relativity}}} (\bibinfo {year} {2024}),\
  \Eprint {https://arxiv.org/abs/2411.16850} {arXiv:2411.16850 [hep-ph]}
  \BibitemShut {NoStop}%
\bibitem [{\citenamefont {Weinberg}\ \emph {et~al.}(2013)\citenamefont
  {Weinberg}, \citenamefont {Arras},\ and\ \citenamefont
  {Burkart}}]{weinberg-etal:13}%
  \BibitemOpen
  \bibfield  {author} {\bibinfo {author} {\bibfnamefont {N.~N.}\ \bibnamefont
  {Weinberg}}, \bibinfo {author} {\bibfnamefont {P.}~\bibnamefont {Arras}},\
  and\ \bibinfo {author} {\bibfnamefont {J.}~\bibnamefont {Burkart}},\
  }\bibfield  {title} {\bibinfo {title} {{An instability due to the nonlinear
  coupling of $p$-modes to $g$-modes: Implications for coalescing neutron star
  binaries}},\ }\href {https://doi.org/10.1088/0004-637X/769/2/121} {\bibfield
  {journal} {\bibinfo  {journal} {Astrophys. J.}\ }\textbf {\bibinfo {volume}
  {769}},\ \bibinfo {pages} {121} (\bibinfo {year} {2013})}\BibitemShut
  {NoStop}%
\bibitem [{\citenamefont {Weinberg}(2016)}]{weinberg:16}%
  \BibitemOpen
  \bibfield  {author} {\bibinfo {author} {\bibfnamefont {N.~N.}\ \bibnamefont
  {Weinberg}},\ }\bibfield  {title} {\bibinfo {title} {{Growth rate of the
  tidal p-mode g-mode instability in coalescing binary neutron stars}},\ }\href
  {https://doi.org/10.3847/0004-637X/819/2/109} {\bibfield  {journal} {\bibinfo
   {journal} {Astrophys. J.}\ }\textbf {\bibinfo {volume} {819}},\ \bibinfo
  {pages} {109} (\bibinfo {year} {2016})}\BibitemShut {NoStop}%
\bibitem [{\citenamefont {Essick}\ \emph {et~al.}(2016)\citenamefont {Essick},
  \citenamefont {Vitale},\ and\ \citenamefont
  {Weinberg}}]{essick-vitale-weinberg:16}%
  \BibitemOpen
  \bibfield  {author} {\bibinfo {author} {\bibfnamefont {R.}~\bibnamefont
  {Essick}}, \bibinfo {author} {\bibfnamefont {S.}~\bibnamefont {Vitale}},\
  and\ \bibinfo {author} {\bibfnamefont {N.~N.}\ \bibnamefont {Weinberg}},\
  }\bibfield  {title} {\bibinfo {title} {{Impact of the tidal
  $p\text{\ensuremath{-}}g$ instability on the gravitational wave signal from
  coalescing binary neutron stars}},\ }\href
  {https://doi.org/10.1103/PhysRevD.94.103012} {\bibfield  {journal} {\bibinfo
  {journal} {Phys. Rev. D}\ }\textbf {\bibinfo {volume} {94}},\ \bibinfo
  {pages} {103012} (\bibinfo {year} {2016})}\BibitemShut {NoStop}%
\bibitem [{\citenamefont {Essick}\ and\ \citenamefont
  {Weinberg}(2018)}]{essick-weinberg:18}%
  \BibitemOpen
  \bibfield  {author} {\bibinfo {author} {\bibfnamefont {R.}~\bibnamefont
  {Essick}}\ and\ \bibinfo {author} {\bibfnamefont {N.~N.}\ \bibnamefont
  {Weinberg}},\ }\href@noop {} {\bibinfo {title} {{A comparison of p-g tidal
  coupling analyses}}} (\bibinfo {year} {2018}),\ \Eprint
  {https://arxiv.org/abs/1809.00264} {arXiv:1809.00264 [astro-ph.HE]}
  \BibitemShut {NoStop}%
\bibitem [{\citenamefont {Abbott}\ \emph {et~al.}(2019)\citenamefont {Abbott},
  \citenamefont {Abbott}, \citenamefont {Abbott}, \citenamefont {Acernese}
  \emph {et~al.}}]{GW170817:19}%
  \BibitemOpen
  \bibfield  {author} {\bibinfo {author} {\bibfnamefont {B.~P.}\ \bibnamefont
  {Abbott}}, \bibinfo {author} {\bibfnamefont {R.}~\bibnamefont {Abbott}},
  \bibinfo {author} {\bibfnamefont {T.~D.}\ \bibnamefont {Abbott}}, \bibinfo
  {author} {\bibfnamefont {F.}~\bibnamefont {Acernese}}, \emph {et~al.}
  (\bibinfo {collaboration} {LIGO Scientific Collaboration and Virgo
  Collaboration}),\ }\bibfield  {title} {\bibinfo {title} {{Constraining the
  $p$-Mode--$g$-Mode Tidal Instability with GW170817}},\ }\href
  {https://doi.org/10.1103/PhysRevLett.122.061104} {\bibfield  {journal}
  {\bibinfo  {journal} {Phys. Rev. Lett.}\ }\textbf {\bibinfo {volume} {122}},\
  \bibinfo {pages} {061104} (\bibinfo {year} {2019})}\BibitemShut {NoStop}%
\bibitem [{\citenamefont {Reyes}\ and\ \citenamefont
  {Brown}(2020)}]{reyes-brown:20}%
  \BibitemOpen
  \bibfield  {author} {\bibinfo {author} {\bibfnamefont {S.}~\bibnamefont
  {Reyes}}\ and\ \bibinfo {author} {\bibfnamefont {D.~A.}\ \bibnamefont
  {Brown}},\ }\bibfield  {title} {\bibinfo {title} {{Constraints on nonlinear
  tides due to p–g mode coupling from the neutron star merger GW170817}},\
  }\href {https://doi.org/10.3847/1538-4357/ab64e8} {\bibfield  {journal}
  {\bibinfo  {journal} {Astrophys. J.}\ }\textbf {\bibinfo {volume} {894}},\
  \bibinfo {pages} {41} (\bibinfo {year} {2020})}\BibitemShut {NoStop}%
\bibitem [{\citenamefont {Kwon}\ \emph {et~al.}(2024)\citenamefont {Kwon},
  \citenamefont {Yu},\ and\ \citenamefont
  {Venumadhav}}]{kwon-yu-venumadhav:24}%
  \BibitemOpen
  \bibfield  {author} {\bibinfo {author} {\bibfnamefont {K.~J.}\ \bibnamefont
  {Kwon}}, \bibinfo {author} {\bibfnamefont {H.}~\bibnamefont {Yu}},\ and\
  \bibinfo {author} {\bibfnamefont {T.}~\bibnamefont {Venumadhav}},\
  }\href@noop {} {\bibinfo {title} {{Resonance locking of anharmonic $g$-modes
  in coalescing neutron star binaries}}} (\bibinfo {year} {2024}),\ \Eprint
  {https://arxiv.org/abs/2410.03831} {arXiv:2410.03831 [gr-qc]} \BibitemShut
  {NoStop}%
\bibitem [{\citenamefont {Pitre}\ and\ \citenamefont
  {Poisson}(2024)}]{pitre-poisson:24}%
  \BibitemOpen
  \bibfield  {author} {\bibinfo {author} {\bibfnamefont {T.}~\bibnamefont
  {Pitre}}\ and\ \bibinfo {author} {\bibfnamefont {E.}~\bibnamefont
  {Poisson}},\ }\bibfield  {title} {\bibinfo {title} {{General relativistic
  dynamical tides in binary inspirals without modes}},\ }\href
  {https://doi.org/10.1103/PhysRevD.109.064004} {\bibfield  {journal} {\bibinfo
   {journal} {Phys. Rev. D}\ }\textbf {\bibinfo {volume} {109}},\ \bibinfo
  {pages} {064004} (\bibinfo {year} {2024})}\BibitemShut {NoStop}%
\bibitem [{\citenamefont {Poisson}(2021)}]{poisson:21a}%
  \BibitemOpen
  \bibfield  {author} {\bibinfo {author} {\bibfnamefont {E.}~\bibnamefont
  {Poisson}},\ }\bibfield  {title} {\bibinfo {title} {{Compact body in a tidal
  environment: New types of relativistic Love numbers, and a post-Newtonian
  operational definition for tidally induced multipole moments}},\ }\href
  {https://doi.org/10.1103/PhysRevD.103.064023} {\bibfield  {journal} {\bibinfo
   {journal} {Phys. Rev. D}\ }\textbf {\bibinfo {volume} {103}},\ \bibinfo
  {pages} {064023} (\bibinfo {year} {2021})}\BibitemShut {NoStop}%
\bibitem [{\citenamefont {Regge}\ and\ \citenamefont
  {Wheeler}(1957)}]{regge-wheeler:57}%
  \BibitemOpen
  \bibfield  {author} {\bibinfo {author} {\bibfnamefont {T.}~\bibnamefont
  {Regge}}\ and\ \bibinfo {author} {\bibfnamefont {J.~A.}\ \bibnamefont
  {Wheeler}},\ }\bibfield  {title} {\bibinfo {title} {{Stability of a
  Schwarzschild singularity}},\ }\href
  {https://doi.org/10.1103/PhysRev.108.1063} {\bibfield  {journal} {\bibinfo
  {journal} {Phys. Rev.}\ }\textbf {\bibinfo {volume} {108}},\ \bibinfo {pages}
  {1063} (\bibinfo {year} {1957})}\BibitemShut {NoStop}%
\bibitem [{\citenamefont {Martel}\ and\ \citenamefont
  {Poisson}(2005)}]{martel-poisson:05}%
  \BibitemOpen
  \bibfield  {author} {\bibinfo {author} {\bibfnamefont {K.}~\bibnamefont
  {Martel}}\ and\ \bibinfo {author} {\bibfnamefont {E.}~\bibnamefont
  {Poisson}},\ }\bibfield  {title} {\bibinfo {title} {{Gravitational
  perturbations of the Schwarzschild spacetime: A practical covariant and
  gauge-invariant formalism}},\ }\href
  {https://doi.org/10.1103/PhysRevD.71.104003} {\bibfield  {journal} {\bibinfo
  {journal} {Phys. Rev. D}\ }\textbf {\bibinfo {volume} {71}},\ \bibinfo
  {pages} {104003} (\bibinfo {year} {2005})}\BibitemShut {NoStop}%
\bibitem [{\citenamefont {Israel}(1966)}]{israel:66}%
  \BibitemOpen
  \bibfield  {author} {\bibinfo {author} {\bibfnamefont {W.}~\bibnamefont
  {Israel}},\ }\bibfield  {title} {\bibinfo {title} {{Singular hypersurfaces
  and thin shells in general relativity}},\ }\href@noop {} {\bibfield
  {journal} {\bibinfo  {journal} {Nuovo Cimento}\ }\textbf {\bibinfo {volume}
  {44}},\ \bibinfo {pages} {1} (\bibinfo {year} {1966})}\BibitemShut {NoStop}%
\bibitem [{\citenamefont {Poisson}\ and\ \citenamefont
  {Will}(2014)}]{poisson-will:14}%
  \BibitemOpen
  \bibfield  {author} {\bibinfo {author} {\bibfnamefont {E.}~\bibnamefont
  {Poisson}}\ and\ \bibinfo {author} {\bibfnamefont {C.~M.}\ \bibnamefont
  {Will}},\ }\href@noop {} {\emph {\bibinfo {title} {Gravity: Newtonian,
  Post-Newtonian, Relativistic}}}\ (\bibinfo  {publisher} {Cambridge University
  Press},\ \bibinfo {address} {Cambridge, England},\ \bibinfo {year}
  {2014})\BibitemShut {NoStop}%
\end{thebibliography}%
\end{document}